\pdfoutput=1
\documentclass[aps,prl,twocolumn,superscriptaddress,floatfix]{revtex4-2}
\usepackage{graphicx,xcolor}
\usepackage{amsmath,amssymb,amsthm}
\usepackage[small,bf,justification=raggedright]{caption}
\usepackage[subrefformat=parens,labelformat=parens, singlelinecheck=false, justification=raggedright]{subcaption}
\usepackage[pdfencoding=auto]{hyperref}
\hypersetup{colorlinks=true,allcolors=black}
\usepackage{physics}
\usepackage[export]{adjustbox}
\usepackage{setspace}
\usepackage{siunitx}

\theoremstyle{definition}

\begin{document}

\title{Linear-optical fusion gate on qudits for efficient generation of entangled qubits}
\title{Linear-optical fusion boosted by high-dimensional entanglement}
\author{Tomohiro Yamazaki}
\email{tomohiro.yamazaki@ntt.com}
\affiliation{NTT Basic Research Laboratories \& NTT Research Center for Theoretical Quantum Information, NTT Corporation, Atsugi, Kanagawa 243-0198, Japan}
\author{Koji Azuma}
\affiliation{NTT Basic Research Laboratories \& NTT Research Center for Theoretical Quantum Information, NTT Corporation, Atsugi, Kanagawa 243-0198, Japan}
\begin{abstract}
We propose a quantum measurement that probabilistically projects a pair of qudits of dimension $d$ onto a Bell state in a two-qubit subspace.
It can be performed using linear-optical circuits with the success probabilities of $1-d^{-1}$ without ancilla photons and $1-d^{-(k+1)}$ with $2(2^{k}-1)$ ancilla photons.
It allows us to entangle two independently-prepared high-dimensional entangled states two-dimensionally with higher probabilities than ones of linear-optical fusion gates on qubits. As an application, we propose a fast quantum repeater protocol with three-qudit GHZ states and quantum memories.
\end{abstract}
\maketitle

\textit{Introduction.---}Photons are crucial careers of quantum information in quantum computation~\cite{Knill_Nature2001,Kok_Rev.Mod.Phys.2007} and communication~\cite{Gisin_Nat.Photonics2007}. 
There, each photon can be used to carry high-dimensional quantum information as a qudit rather than a qubit by using its degrees of freedom such as path~\cite{Carine_Optica2020}, time-bin~\cite{Humphreys_Phys.Rev.Lett.2013}, frequency-bin~\cite{Lukens_Optica2017,Cui_Phys.Rev.Lett.2020}, time-frequency~\cite{Brecht_Phys.Rev.X2015,Fabre_Phys.Rev.A2022}, frequency-comb~\cite{Fabre_Phys.Rev.A2020a,Yamazaki_Phys.Rev.Lett.2023}, and orbital angular momentum~\cite{Vaziri_Phys.Rev.Lett.2002,Erhard_LightSci.Appl.2018}.
A qudit can not only carry $\log_2{d}$ qubits of information but also show higher noise resistance than a qubit~\cite{Collins_Phys.Rev.Lett.2002,Cerf_Phys.Rev.Lett.2002,Zhu_AVSQuantumSci.2021}.
Besides, increasing the dimension of a photonic qudit seems to be experimentally less demanding than increasing the number of photonic qubits because the former merely requires more photonic modes while the latter requires more photons.

Despite such benefits, it is challenging to show the superiority of photonic qudits over photonic qubits in applications, except for ones for small-scale quantum information processing (QIP) such as quantum key distribution~\cite{Cozzolino_Adv.QuantumTechnol.2019}.
In applications for large-scale QIP such as linear-optical quantum computation~\cite{Browne_Phys.Rev.Lett.2005,Kieling_Phys.Rev.Lett.2007,Gimeno-Segovia_Phys.Rev.Lett.2015,Bartolucci_Nat.Commun.2023,Sahay_Phys.Rev.Lett.2023} and repeaters~\cite{Azuma_Nat.Commun.2015,Ewert_Phys.Rev.Lett.2016},
it is an essential step to generate larger entangled states from small entangled states with probabilistic entangling measurements, called fusion gates~\cite{Browne_Phys.Rev.Lett.2005}.
While linear-optical fusion gates for qubits have the success probability of $50\%$ without ancilla photons~\cite{Weinfurter_EPL1994,Braunstein_Phys.Rev.A1995,Calsamiglia_Appl.Phys.B2001}, linear-optical fusion gates for qudits require at least $d-2$ ancilla photons to succeed~\cite{Calsamiglia_Phys.Rev.A2002}, and the success probabilities of known schemes are up to $O(d^{-2})$~\cite{Luo_Phys.Rev.Lett.2019,Zhang_Phys.Rev.A2019a, Bharos_2024}.
Although a scheme for high-dimensional linear-optical quantum computation has recently been proposed~\cite{Paesani_Phys.Rev.Lett.2021}, the number of photons required to generate a large entangled state of qudits is too large to be offset by the increase in information per photon, compared to the case of qubits.

Another approach to leveraging photonic qudits is to use them for assisting QIP based on qubits, rather than qudits.
Recent quantum computation protocols with photonic qubits~\cite{Gimeno-Segovia_Phys.Rev.Lett.2015,Bartolucci_Nat.Commun.2023,Sahay_Phys.Rev.Lett.2023} usually assume fusion gates with a high success probability.
Such fusion gates can be realized by using ancilla photons~\cite{Grice_Phys.Rev.A2011,Zaidi_Phys.Rev.Lett.2013,Ewert_Phys.Rev.Lett.2014,Olivo_Phys.Rev.A2018,Bartolucci_2021b}, but the size of entangled photons required as the ancilla increases linearly with $N$ to achieve the success probability of $1-N^{-1}$.
On the other hand, it is known that Bell states defined in a two-qubit subsystem of two ququarts~\footnote{A ququart is a qudit of dimension $d=4$. The original scheme, called hyperentanglement-assisted Bell-state analysis, is proposed in terms of hyperentanglement, where each photon forms two qubits with two degrees of freedom of the photon.} can be deterministically discriminated with the help of additional entanglement in the other two-qubit subsystem~~\cite{Walborn_Phys.Rev.A2003,Schuck_Phys.Rev.Lett.2006}.
This boosting up of the success probability enabled by the use of qudits is highly appealing.
However, its application is limited to small-scale QIP~\cite{Barreiro_Nat.Phys.2008} because this scheme needs the prior entanglement between the inputs.
Therefore, photonic qudits still seem to lack compelling applications for large-scale QIP.

In this paper, we propose a new type of quantum measurements on qudits, we call \textit{pairwise fusion gate} (PFG).
The PFG includes projections onto one of ``pairwise'' Bell states $\ket*{\psi_{i,j,\pm}} = \sqrt{2^{-1}}(\ket{i}\ket{j} \pm \ket{j}\ket{i})$ and $\ket*{\phi_{i,j,\pm}}= \sqrt{2^{-1}}(\ket{i}\ket{i} \pm\ket{j}\ket{j})$ defined on a two-qubit subspace specified by a pair of different integers $i,j \in \mathbb{Z}_d=\{0,\cdots, d-1\}$.
The success probability $p_\text{f}$ of the PFG is defined as the average probability of one of such projections being applied over arbitrary inputs.
We present a linear-optical circuit (LOC) without ancilla photons, which works as a PFG with $p_\text{f}=1-d^{-1}$.
In addition, we present LOCs using ancilla photons that work as PFGs with higher success probabilities, which scale as $p_\text{f}=1-d^{-(k+1)}$ for $2(2^k-1)$ ancilla photons for an integer $k\geq 1$.
By applying a PFG to two independently-prepared high-dimensional entangled states, we can entangle them two-dimensionally with the probability of $p_\text{f}$, which is higher than one of two-dimensional analogues.
As an illustrating example of applications of PFGs, we propose a quantum repeater (QR) protocol using photonic three-qudit GHZ states and quantum memories.
Thanks to the high success probability of the PFG, our protocol greatly outperforms standard QR protocols based on linear-optical fusion gates on qubits and quantum memories.

\textit{Construction of PFG.---}We introduce a set of LOCs, composed of linear-optical elements, photon-number-resolving detectors, and optionally ancilla photons, that functions as the PFG. 
Any time evolution of Fock states induced by the interferometric part of an LOC, consisting only of linear-optical elements, is represented as a transformation $a_i^\dag \rightarrow \sum_{j=0}^{m-1} a^\dag_j u_{ji}$ of creation operators $\{a_i^\dag\}_{i=0,\dots,m-1}$ with a unitary matrix $U=[u_{ji}]$, which is called \textit{transfer matrix}.
Hereafter, the interferometric part of an LOC is specified by its corresponding transfer matrix.
Note that, conversely, for any unitary matrix $U$, the time evolution represented by $a_i^\dag \rightarrow \sum_{j=0}^{m-1} a^\dag_j u_{ji}$ can be implemented by combining linear-optical elements.
For instance, an LOC specified by the $2 \times 2$ Hadamard matrix $H$ is realized by a 50:50 beam splitter.

We need to encode qudits in the Fock space for QIP. 
Here, we use the encoding called \textit{$d$-rail encoding}, where a qudit of dimension $d$ is defined with a photon on $d$ modes as
\begin{equation}
    \ket{i} \leftrightarrow \ket*{\nu_i} \equiv \ket*{\underbrace{0,\cdots, 0}_{i}, 1, \underbrace{0, \cdots, 0}_{d-(i+1)}}
\end{equation}
for $i \in \mathbb{Z}_d$, where the left-hand (right-hand) side is a state of a qudit (photon).
In the case of multiple qudits, a basis state $\ket{i}$ of the $n$th ($n \geq 1$) qudit is associated with a single-photon state $\ket{\nu_i}$ defined on the $(n-1)d$th through $(nd-1)$th modes, e.g., $\ket{i}\ket{j}\leftrightarrow\ket*{\nu_{i}}\ket*{\nu_{j}}=a^\dag_i a^\dag_{d+j}\ket*{\text{vac}}$ for $n=2$.
For a state $\ket*{\Psi}$ of qudits, we use $\ket*{\Psi^{\nu}}$ to represent the corresponding state in the Fock space, where basis states $\ket{i}$ for $\ket*{\Psi}$ are replaced with Fock states $\ket*{\nu_i}$ for $\ket*{\Psi^{\nu}}$.
We consider to input two photonic qudits into an LOC that outputs only a measurement pattern $(n_0,n_1,\cdots)$ from the detectors (that is, there is no output photon).
Then, the LOC works as a quantum measurement on two qudits, which is what we need to identify.

\begin{figure}[tpb]
\begin{minipage}[b]{0.48\linewidth}
\centering
\subcaption{}
\vspace{-10pt}
\includegraphics[scale=0.7, trim=0 0 0 0, clip]{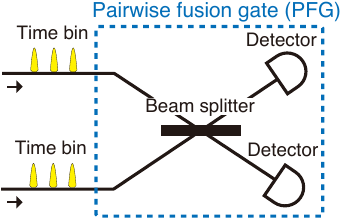}
\label{fig:timebin_setup}
\vspace{-10pt}
\end{minipage}
\begin{minipage}[b]{0.49\linewidth}
\centering
\subcaption{}
\vspace{-10pt}
\includegraphics[scale=0.53, trim=0 0 0 0, clip]{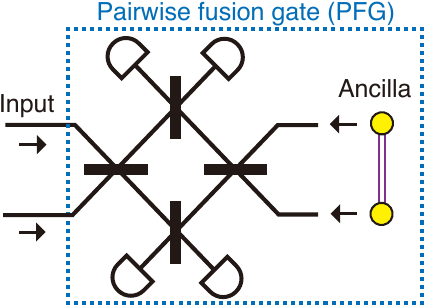} 
\label{fig:timebin_setup_ancilla}
\end{minipage}
\caption{Experimental setups of pairwise fusion gates (PFGs) for time-bin qudits (a) without ancilla photons and (b) with two photons in a high-dimensional Bell state.}
\end{figure}
First, we consider the LOC consisting of the interferometric part specified by $H \otimes I_d$ and $2d$ detectors that is depicted, for instance, as in Fig.~\ref{fig:timebin_setup} assuming time-bin qudits.
In the time-bin encoding, a basis state $\ket{i}$ of each qudit for $i\in \mathbb{Z}_d$ is associated with a single photon occupying the $i$th time bin on each path.
The experimental setup in Fig.~\ref{fig:timebin_setup} consists of a time-independent 50:50 beam splitter between two path modes and two time-resolving photon detectors.
Since the beam splitter does not change the temporal modes of photons, the total number of photons in the $i$th time slot across all the two path modes, represented by $N_i=a_{i}^\dag a_{i}+a_{d+i}^\dag a_{d+i}$, is preserved until the photons reach the detectors.

We classify the possible measurement patterns from the detectors according to $(N_i)_{i\in \mathbb{Z}_d}$.
There are two cases: (i) $N_i=2$ for $i \in \mathbb{Z}_d$, two photons are detected at the same time slot indexed by $i$, and (ii) $N_i=1$ and $N_j=1$ for different $i, j \in \mathbb{Z}_d$, two photons are detected at the different time slots indexed by $i$ and $j$, respectively.
We consider inputting basis states $\{\ket{i}\ket{j} \mid i,j \in \mathbb{Z}_d\}$ of two qudits into the LOC.
For any $i \in \mathbb{Z}_d$, $\ket{i}\ket{i}$ is the unique input state giving a measurement pattern satisfying $N_i=2$ among the basis states because $N_i$ is a conserved quantity.
Thus, any measurement pattern satisfying $N_i=2$ corresponds to the application of the Kraus operator $\bra{i}\bra{i}$ to the input qudits.
On the other hand, for any different $i, j \in \mathbb{Z}_d$, both $\ket{i}\ket{j}$ and $\ket{j}\ket{i}$  can give a measurement pattern satisfying $N_i=1$ and $N_j=1$.
Thus, the Kraus operators corresponding to a measurement pattern satisfying $N_i=1$ and $N_j=1$ are spanned by $\bra{i}\bra{j}$ and $\bra{j}\bra{i}$.
To determine the corresponding Kraus operators, we calculate the time evolution of input states $\ket*{\psi_{i,j,+}}$ and $\ket*{\psi_{i,j,-}}$ as
\begin{align}
    a^\dag_i a^\dag_{d+j}+ a^\dag_j a^\dag_{d+i} &\rightarrow a^\dag_i a^\dag_{j} - a^\dag_{d+i} a^\dag_{d+j}, \label{eq:psi+}\\
    a^\dag_i a^\dag_{d+j}- a^\dag_j a^\dag_{d+i} &\rightarrow a^\dag_{j} a^\dag_{d+i} - a^\dag_i a^\dag_{d+j} \label{eq:psi-},
\end{align}
respectively.
Thus, $\ket*{\psi_{i,j,+}}$ and $\ket*{\psi_{i,j,-}}$ always give different measurement patterns, which means that each measurement pattern corresponds to the application of Kraus operator $\bra*{\psi_{i,j,+}}$ or $\bra*{\psi_{i,j,-}}$.
Therefore, we identify that the LOC corresponds to the quantum measurement represented by the set of Kraus operators,
\begin{equation}\label{eq:no_ancilla}
    \{\bra{i}\bra{i}, \bra*{\psi_{i,j,+}}, \bra*{\psi_{i,j,-}}\}
\end{equation}
for $i,j\in \mathbb{Z}_d$ with $i < j$, which is a PFG with $p_\text{f}=1-d^{-1}$.

Next, we consider LOCs using ancilla photons to obtain PFGs with higher success probabilities.
For any integer $k\geq 0$, we propose an LOC that consists of the interferometric part specified by $H^{\otimes k+1} \otimes I_d$, $2^{k+1} d$ detectors, and ancilla photons in state $\bigotimes_{q=1}^{k} \ket*{\text{GHZ}^{\nu}(2^q)}$, where $\ket*{\text{GHZ}(2^q)} = \sqrt{d^{-1}} \sum_{i\in \mathbb{Z}_d} \ket{i}^{\otimes 2^q}$ is a $2^q$-qudit GHZ state.
This scheme is reduced to a known scheme proposed by Grice~\cite{Grice_Phys.Rev.A2011} for $d=2$.
In the Supplemental Material, we analyze these LOCs with the \textit{bosonic stabilizer formalism}, which is a group-theoretic framework to analyze LOCs, recently proposed in~\cite{Yamazaki_2023}.
As a result, we find that the considered LOC corresponds to the quantum measurement on two qudits, which is represented by the set of Kraus operators,
\begin{equation}\label{eq:n_ancilla}
    \{\sqrt{d^{-k}}\bra{i}\bra{i}, \sqrt{c_k}\bra*{\phi_{i,j,+}}, \sqrt{c_k}\bra*{\phi_{i,j,-}}, \bra*{\psi_{i,j,+}}, \bra*{\psi_{i,j,-}}\}
\end{equation}
for $i,j\in \mathbb{Z}_d$ with $i < j$, where $c_k = \sum_{p=1}^{k}d^{-p}$.
This is a PFG with $p_\text{f} = 1-d^{-(k+1)}$, which is dramatically increased even for small $k$, depending on $d$.

To understand why the use of ancilla photons makes the success probability higher, we consider the case of $k=1$ using the time-bin encoding.
The corresponding experimental setup for time-bin qudits is shown in Fig.~\ref{fig:timebin_setup_ancilla}, where the ancilla photons are in a $d$-dimensional Bell state $\ket*{\phi^{\nu}}$, where $\ket*{\phi}=\ket*{\text{GHZ}(2)}=\sqrt{d^{-1}}\sum_{i\in \mathbb{Z}_d}\ket{i}\ket{i}$.
As in the case of no ancilla, we classify the measurement patterns from the detectors according to conserved quantities $N_i= \sum_{k=0}^3 a_{dk+i}^\dag a_{dk+i}$ for $i \in \mathbb{Z}_d$.
Note that two photons in ancilla state $\ket*{\phi^{\nu}}$ are always in a same time slot and contribute to increasing $N_i$ by two for $i \in \mathbb{Z}_d$.
Thus, there are four cases: (i) $N_i=4$, (ii) $N_i=3$ and $N_j=1$, (iii) $N_i=2$ and $N_j=2$, and (iv) $N_i=1$, $N_j=1$, and $N_k=2$ for different $i, j, k \in  \mathbb{Z}_d$.
We consider inputting basis states $\{\ket{i}\ket{j} \mid i,j \in \mathbb{Z}_d\}$ of two qudits into the LOC.
In case (i), $\ket{i}\ket{i}$ is the unique input state that can give a measurement pattern satisfying $N_i=4$ among the basis states.
Similarly, $\ket{i}\ket{j}$ and $\ket{j}\ket{i}$ are unique input states resulting in case (ii) or case (iv), and $\ket{i}\ket{i}$ and $\ket{j}\ket{j}$ are unique input states falling into case (iii) among the basis states.
While case (i) clearly corresponds to a projection onto $\ket{i}\ket{i}$, we can ensure that cases (ii) and (iv) (case (iii)) correspond to a projection onto $\ket*{\psi_{i,j,\pm}}$ ($\ket*{\phi_{i,j,\pm}}$).
Thus, unlike in the case of no ancilla, this LOC has a chance to succeed as a PFG even for input states $\{\ket{i}\ket{i}\}$ by obtaining case (iii) rather than case (i), which occurs with the probability of $1-d^{-1}$.
In total, this LOC works as a PFG with $p_\text{f} = 1-d^{-2}$.

\begin{figure}[tpb]
    \begin{minipage}[b]{\linewidth}
        \centering
        \subcaption{}
        \vspace{-10pt}
        \hspace{25pt}
        \includegraphics[scale=0.65, trim=0 0 0 0, clip]{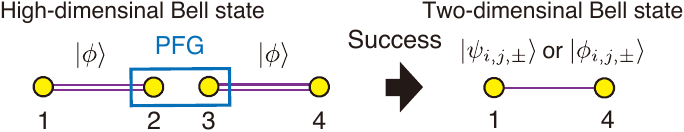} 
        \label{fig:fusion_scheme1}
    \end{minipage}\\
    \begin{minipage}[b]{\linewidth}
        \centering
        \subcaption{}
        \vspace{-10pt}
        \hspace{9pt}
        \includegraphics[scale=0.65, trim=0 0 0 0, clip]{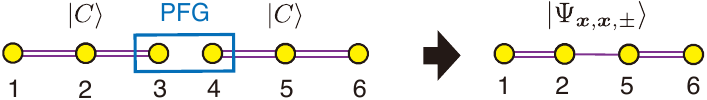}
        \label{fig:fusion_scheme2}
    \end{minipage}
    \begin{minipage}[b]{\linewidth}
        \centering
        \subcaption{}
        \vspace{-10pt}
        \hspace{9pt}
        \includegraphics[scale=0.65, trim=0 0 0 0, clip]{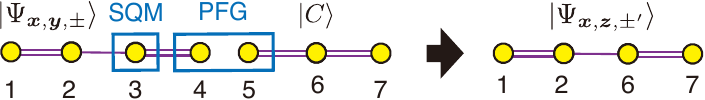}
        \label{fig:fusion_scheme3}
    \end{minipage}
    \begin{minipage}[b]{\linewidth}
        \centering
        \subcaption{}
        \includegraphics[scale=0.65, trim=0 0 0 0, clip]{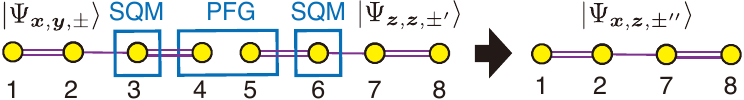}
        \label{fig:fusion_scheme4}
    \end{minipage}
    \caption{State conversions with PFGs and single-qudit measurements (SQMs) (a) from $\ket*{\phi}^{\otimes 2}$ to $\ket*{\psi_{i,j,\pm}}$ or $\ket*{\phi_{i,j,\pm}}$, (b) from $\ket*{C}^{\otimes 2}$ to $\ket{\Psi_{\vb*{x},\vb*{x},\pm}}$, (c) from $\ket*{\Psi_{\vb*{x},\vb*{y},\pm}}\ket*{C}$ to $\ket*{\Psi_{\vb*{x},\vb*{z},\pm'}}$, and (d) from $\ket*{\Psi_{\vb*{x},\vb*{y},\pm}}\ket*{\Psi_{\vb*{z},\vb*{z},\pm'}}$ to $\ket{\Psi_{\vb*{x},\vb*{z},\pm''}}$, where $\pm$, $\pm'$, and $\pm''$ can take different signs. The SQMs to be applied depend on the measurement outcomes of the PFGs. The success probabilities of these conversions are $p_\text{f}$.}
\end{figure}
\textit{Application of PFG.---}Here, we consider how to use PFGs for generating entangled qubits with high success probabilities.
The simplest example is to apply a PFG to halves of two pairs of $d$-dimensional Bell state $\ket*{\phi}^{\otimes 2}$, as shown in Fig.~\ref{fig:fusion_scheme1}.
Then, we obtain a pairwise Bell state $\ket*{\psi_{i,j,\pm}}$ or $\ket*{\phi_{i,j,\pm}}$ for different $i, j \in \mathbb{Z}_d$ with the total probability of $p_\text{f}$.
By discarding all modes except for the $i$th and $j$th modes, this state can be reduced to a Bell state defined on two qubits.
Generalizing this process, for example, we can generate a $2(n-1)$-qubit GHZ state from two $n$-qudit GHZ state $\ket*{\text{GHZ}(n)}^{\otimes 2}$ with the probability of $p_\text{f}$.
In these applications, the output state of a PFG is immediately reduced to a state of qubits, which implies that we cannot enjoy the boosted success probability again by applying other PFGs to the output state.
However, that is not always the case.
In the following, we present a nontrivial example in which multiples PFGs can be applied while keeping the boosted success probability, which we call \textit{boosted entanglement swapping} (BES).
The final output state of a series of BESs is just a Bell state of qubits, which can be applied for quantum communication.

First, we consider a three-qudit state $\ket*{C}=d^{-1}\sum_{i,j \in \mathbb{Z}_d}\ket{i,i+j,j}$ as an initial state, where $\ket{i}$ for $i\geq d$ means $\ket{i \pmod{d}}$.
This is equivalent to a three-qudit GHZ state under local unitary transformations.
As shown in Fig~\ref{fig:fusion_scheme2}, a successful application of a PFG to two copies of $\ket*{C}$ outputs the state in $\ket*{\Psi_{\vb*{x},\vb*{x},\pm}}$, where $\ket*{\Psi_{\vb*{x},\vb*{y},\pm}}$ is defined by
\begin{equation}\label{intermediate}
    \frac{1}{d\sqrt{2}}\sum_{i,j \in \mathbb{Z}_d}(\ket{i,i+x_0,j+y_1,j} \pm \ket{i,i+x_1,j+y_0,j})
\end{equation}
for $\vb*{x}=(x_0, x_1)$ and $\vb*{y}=(y_0, y_1) \in \mathbb{Z}_d^2$ with $x_0 \neq x_1$ and  $y_0 \neq y_1$.
Although $\ket*{\Psi_{\vb*{x},\vb*{y},\pm}}_{1234}$ is two-dimensionally entangled in bipartition of $12|34$, it it still $d$-dimensionally entangled in bipartition of $1|234$ and $123|4$. 

\begin{figure*}[tbp]
\centering
\hspace{-30pt}
\includegraphics[scale=0.5, trim=0 0 0 0, clip]{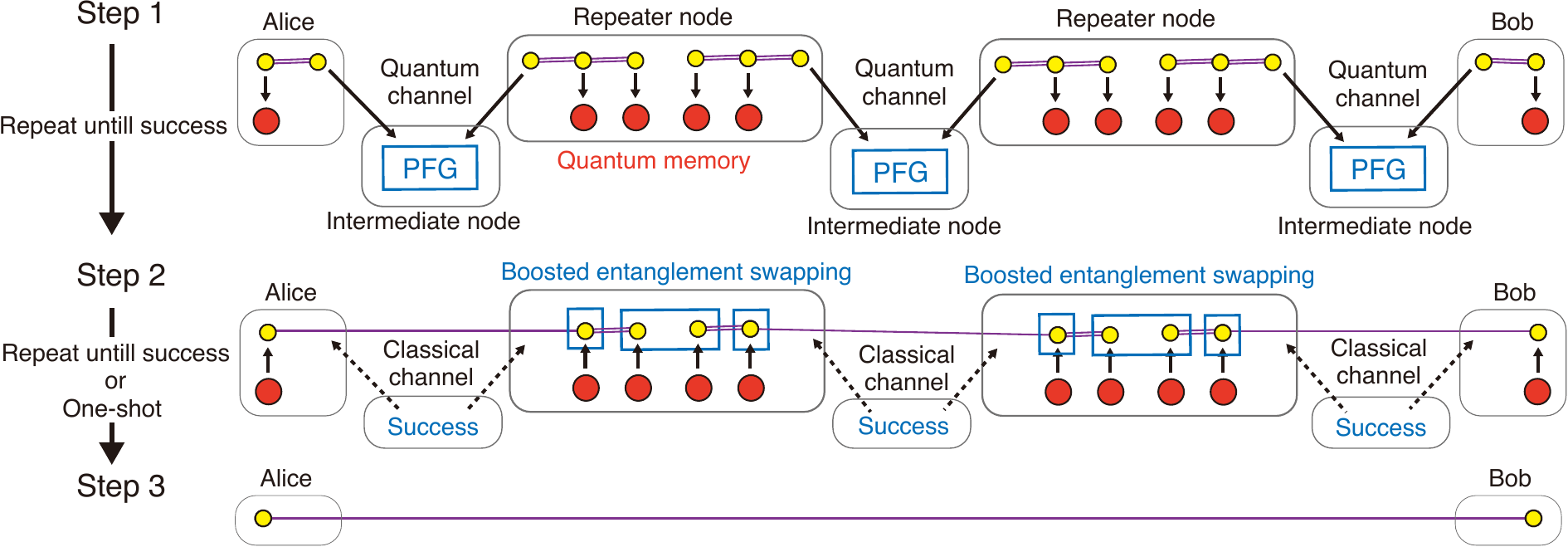}
\caption{A quantum repeater protocol with PFGs and quantum memories. Each repeater node (RN) prepares two copies of photonic three-qudit state $\ket*{C}$, stores two photons of each $\ket*{C}$ in two quantum memories, and send the other photon to each adjacent intermediate node (IN). Each IN performs a PFG on received two photons and, if the PFG succeeds, informs the adjacent RNs of the measurement outcome of the PFG. Each RN and IN repeat this step until success. After  receiving the measurement outcomes from both of the adjacent INs, each RN retrieves the remaining photons from the quantum memories and performs a boosted entanglement swapping (BES) on them.
Depending on the protocol, each BES is repeated until success with a sequential order, or performed independently and almost simultaneously. Once all the BESs succeed, a Bell state of qubits is shared between Alice and Bob.}\label{fig:repeater_protocol}
\end{figure*}

Next, we consider a conversion from $\ket*{\Psi_{\vb*{x},\vb*{y},\pm}}\ket*{C}$ to $\ket*{\Psi_{\vb*{x},\vb*{z},\pm'}}$ as shown in Fig.~\ref{fig:fusion_scheme3}, where $\pm$ and $\pm'$ can take different signs.
We first apply a PFG to the qudits $4$ and $5$ in $\ket*{\Psi_{\vb*{x},\vb*{y},\pm}}_{1234}\ket*{C}_{567}$.
When the PFG works as, for instance, a projection $\bra*{\psi_{z_0,z_1,+}}_{45}$,
the resulting state is 
\begin{equation}\label{eq:PsiC_result}
    \frac{1}{2d}\sum_{i,j \in \mathbb{Z}_d}\sum_{a,b \in \{0,1\}}  (\pm1)^{a} \ket{i,i+x_{a},z_b+y_{a \oplus 1},z_{b\oplus 1}+j,j}_{12367},
\end{equation}
where $\oplus$ denotes addition module $2$.
The qudit $3$ in Eq.~\eqref{eq:PsiC_result} is in one of the four possible states $\{\ket*{z_0+y_1},\ket*{z_0+y_0},\ket*{z_1+y_1},\ket*{z_1+y_0}\}$, but
two of these states may coincide when conditions (i) $y_0+z_0=y_1+z_1$ and/or (ii) $y_0+z_1=y_1+z_0$ are satisfied.
Thus, depending on the measurement outcome, the qudit $3$ effectively becomes a qubit, qutrit, or ququart (where the qubit case can be removed by taking odd $d$~\footnote{Each condition is written as $\Delta_{y} \pm \Delta_{z}=0 \pmod{d}$ where $\Delta_{y}=y_1-y_0$ and $\Delta_{z}=z_1-z_0$. The both conditions are satisfied iff $2\Delta_y = 2\Delta_z = 0 \pmod{d}$, that is, $\Delta_y = \Delta_z = d/2$, which imposes $d$ to be even}).
To remove this degree of freedom, we perform a single-qudit measurement $\mathcal{M}(\vb*{y},\vb*{z})$ on the qudit $3$.
Here, $\mathcal{M}(\vb*{y},\vb*{z})$ is defined as a measurement including, at least, $\{ \sqrt{2^{-1}}(\bra*{y_0+z_0}\pm\bra*{y_1+z_1}), \sqrt{2^{-1}}(\bra*{y_0+z_1}\pm\bra*{y_1+z_0})\}$ as its Kraus operators when neither condition (i) nor condition (ii) is not satisfied, and the Kraus operator $\sqrt{2^{-1}}(\bra*{y_0+z_0}\pm\bra*{y_1+z_1})$ ($\sqrt{2^{-1}}(\bra*{y_0+z_1}\pm\bra*{y_1+z_0})$) is replaced with $\bra*{y_0+z_0}$ ($\bra*{y_0+z_1}$) when condition (i) (condition (ii)) is satisfied.
Applying $\mathcal{M}(\vb*{y},\vb*{z})$ to the state in Eq.~\eqref{eq:PsiC_result}, we can obtain $\ket*{\Psi_{\vb*{x},\vb*{z},\pm'}}_{1267}$, where $\pm'$ is determined by the measurement outcomes of the PFG and $\mathcal{M}(\vb*{y},\vb*{z})$ in addition to $\pm$ of initial state $\ket*{\Psi_{\vb*{x},\vb*{y},\pm}}$.

Finally, we consider the BES, which converts $\ket*{\Psi_{\vb*{x},\vb*{y},\pm}}\ket*{\Psi_{\vb*{z},\vb*{z},\pm'}}$ to $\ket*{\Psi_{\vb*{x},\vb*{z},\pm''}}$ as shown in Fig.~\ref{fig:fusion_scheme4}.
A BES consists of a PFG and two single-qudit measurements $\mathcal{M}(\vb*{y},\vb*{w})$ and $\mathcal{M}(\vb*{w},\vb*{z})$, where $\vb*{w}$ is determined by the outcome of the PFG.
Since $\ket*{\Psi_{\vb*{z},\vb*{z},\pm'}}$ can be considered as an output state after applying a PFG to $\ket*{C}^{\otimes 2}$ as in Fig.~\ref{fig:fusion_scheme2}, a BES can be understood as having applied the conversion from $\ket*{\Psi_{\vb*{x},\vb*{y},\pm}}\ket*{C}$ to $\ket*{\Psi_{\vb*{x},\vb*{z},\pm'}}$ as in Fig.~\ref{fig:fusion_scheme3} twice.
Thus, the BES also succeeds with the probability of $p_\text{f}$.

The BES can be applied to a QR protocol as shown in Fig.~\ref{fig:repeater_protocol} (see the Supplemental Material for more details).
It consists of quantum memories for photonic qudits and GHZ states of photonic qudits, which have been demonstrated experimentally with several physical systems~\cite{Zhou_Phys.Rev.Lett.2015,Ding_LightSci.Appl.2016,Zhang_Nat.Commun.2016,Tiranov_Phys.Rev.A2017,Dong_Phys.Rev.Lett.2023} and can be generated even with LOCs~\cite{Paesani_Phys.Rev.Lett.2021,Xing_Opt.Express2023,Chin_2024}, respectively.
In the case that the BESs are performed in sequence with classical communication among repeater nodes, the protocol is classified to so-called first-generation QRs~\cite{Munro_IEEEJ.Sel.Top.QuantumElectron.2015,Muralidharan_Sci.Rep.2016,Azuma_Rev.Mod.Phys.2023}, similar to standard protocols with linear optics and quantum memories, such as atomic ensembles~\cite{Briegel_Phys.Rev.Lett.1998a,Duan_Nature2001,Sangouard_Rev.Mod.Phys.2011}.
Although first-generation QRs can work even with entanglement swapping with a low success probability, the scaling of its entanglement distribution time with respect to the total distance depends on the success probability of entanglement swapping.
In particular, the scaling becomes linear only when the success probability of entanglement swapping is higher than a threshold around $3/4$ (see the Supplemental Material), which can be achieved in our protocol but not in the standard first-generation QR protocols. 
In fact, we numerically show that our protocol can substantially outperform the conventional protocol thanks to the high success probability of the BES (see the Supplemental Material).
On the other hand, when the BES in each repeater node is performed independently, our protocol can be classified to 
second-generation QRs~\cite{Munro_IEEEJ.Sel.Top.QuantumElectron.2015,Muralidharan_Sci.Rep.2016,Azuma_Rev.Mod.Phys.2023}, which do not need classical communication between non-adjacent nodes and thereby can be faster and relax the requirement for coherence time of quantum memories.
The second-generation QRs~\cite{Jiang_Phys.Rev.A2009,Munro_Nat.Photonics2010,Li_NewJ.Phys.2013} normally require matter qubits equipped with (near-)deterministic entangling gates, such as trapped ions~\cite{Sangouard_Phys.Rev.A2009,Asadi_Quantum2018,Asadi_QuantumSci.Technol.2020} because the entanglement swapping has to be almost deterministic.
In contrast, as demonstrated in the numerical simulations, our protocol has the potential to achieve the second-generation QRs without such matter qubits.

\textit{Conclusion.---}We introduced the PFG, which projects two input qudits onto a Bell state in a two-qubit subspace, and proposed the LOCs that work as PFGs with high success probabilities.
We also discussed applications of PFGs, especially, for QRs.
The proposed QR protocol itself is worthwhile of further analysis because it provides a new avenue for accomplishing fast long-distance quantum communication.
On the other hand, the protocol is nothing other than the generation of a Bell state of qubits using PFGs, and the generation of other important states, such as cluster states, using PFGs remains unexplored.
Further theoretical studies on lower-dimensional entanglement in qudit system seems to be necessary to elucidate this.
This work presents a new use of entangled photonic qudits for qubit-based QIP, which we expect will motivate the development of entangled-photon generators~\cite{Lindner_Phys.Rev.Lett.2009,Schwartz_Science2016} to extend them from qubits to qudits, as discussed in~\cite{Bell_NewJ.Phys.2022,Raissi_PRXQuantum2024}.

We thank William J. Munro for helpful discussion.

\bibliography{hdbell.bib}

\begin{thebibliography}{66}%
\makeatletter
\providecommand \@ifxundefined [1]{%
 \@ifx{#1\undefined}
}%
\providecommand \@ifnum [1]{%
 \ifnum #1\expandafter \@firstoftwo
 \else \expandafter \@secondoftwo
 \fi
}%
\providecommand \@ifx [1]{%
 \ifx #1\expandafter \@firstoftwo
 \else \expandafter \@secondoftwo
 \fi
}%
\providecommand \natexlab [1]{#1}%
\providecommand \enquote  [1]{``#1''}%
\providecommand \bibnamefont  [1]{#1}%
\providecommand \bibfnamefont [1]{#1}%
\providecommand \citenamefont [1]{#1}%
\providecommand \href@noop [0]{\@secondoftwo}%
\providecommand \href [0]{\begingroup \@sanitize@url \@href}%
\providecommand \@href[1]{\@@startlink{#1}\@@href}%
\providecommand \@@href[1]{\endgroup#1\@@endlink}%
\providecommand \@sanitize@url [0]{\catcode `\\12\catcode `\$12\catcode `\&12\catcode `\#12\catcode `\^12\catcode `\_12\catcode `\%12\relax}%
\providecommand \@@startlink[1]{}%
\providecommand \@@endlink[0]{}%
\providecommand \url  [0]{\begingroup\@sanitize@url \@url }%
\providecommand \@url [1]{\endgroup\@href {#1}{\urlprefix }}%
\providecommand \urlprefix  [0]{URL }%
\providecommand \Eprint [0]{\href }%
\providecommand \doibase [0]{https://doi.org/}%
\providecommand \selectlanguage [0]{\@gobble}%
\providecommand \bibinfo  [0]{\@secondoftwo}%
\providecommand \bibfield  [0]{\@secondoftwo}%
\providecommand \translation [1]{[#1]}%
\providecommand \BibitemOpen [0]{}%
\providecommand \bibitemStop [0]{}%
\providecommand \bibitemNoStop [0]{.\EOS\space}%
\providecommand \EOS [0]{\spacefactor3000\relax}%
\providecommand \BibitemShut  [1]{\csname bibitem#1\endcsname}%
\let\auto@bib@innerbib\@empty
\bibitem [{\citenamefont {Knill}\ \emph {et~al.}(2001)\citenamefont {Knill}, \citenamefont {Laflamme},\ and\ \citenamefont {Milburn}}]{Knill_Nature2001}%
  \BibitemOpen
  \bibfield  {author} {\bibinfo {author} {\bibfnamefont {E.}~\bibnamefont {Knill}}, \bibinfo {author} {\bibfnamefont {R.}~\bibnamefont {Laflamme}},\ and\ \bibinfo {author} {\bibfnamefont {G.~J.}\ \bibnamefont {Milburn}},\ }\bibfield  {title} {\bibinfo {title} {A scheme for efficient quantum computation with linear optics},\ }\href {https://doi.org/10.1038/35051009} {\bibfield  {journal} {\bibinfo  {journal} {Nature}\ }\textbf {\bibinfo {volume} {409}},\ \bibinfo {pages} {46} (\bibinfo {year} {2001})}\BibitemShut {NoStop}%
\bibitem [{\citenamefont {Kok}\ \emph {et~al.}(2007)\citenamefont {Kok}, \citenamefont {Munro}, \citenamefont {Nemoto}, \citenamefont {Ralph}, \citenamefont {Dowling},\ and\ \citenamefont {Milburn}}]{Kok_Rev.Mod.Phys.2007}%
  \BibitemOpen
  \bibfield  {author} {\bibinfo {author} {\bibfnamefont {P.}~\bibnamefont {Kok}}, \bibinfo {author} {\bibfnamefont {W.~J.}\ \bibnamefont {Munro}}, \bibinfo {author} {\bibfnamefont {K.}~\bibnamefont {Nemoto}}, \bibinfo {author} {\bibfnamefont {T.~C.}\ \bibnamefont {Ralph}}, \bibinfo {author} {\bibfnamefont {J.~P.}\ \bibnamefont {Dowling}},\ and\ \bibinfo {author} {\bibfnamefont {G.~J.}\ \bibnamefont {Milburn}},\ }\bibfield  {title} {\bibinfo {title} {Linear optical quantum computing with photonic qubits},\ }\href {https://doi.org/10.1103/RevModPhys.79.135} {\bibfield  {journal} {\bibinfo  {journal} {Rev. Mod. Phys.}\ }\textbf {\bibinfo {volume} {79}},\ \bibinfo {pages} {135} (\bibinfo {year} {2007})}\BibitemShut {NoStop}%
\bibitem [{\citenamefont {Gisin}\ and\ \citenamefont {Thew}(2007)}]{Gisin_Nat.Photonics2007}%
  \BibitemOpen
  \bibfield  {author} {\bibinfo {author} {\bibfnamefont {N.}~\bibnamefont {Gisin}}\ and\ \bibinfo {author} {\bibfnamefont {R.}~\bibnamefont {Thew}},\ }\bibfield  {title} {\bibinfo {title} {Quantum communication},\ }\href {https://doi.org/10.1038/nphoton.2007.22} {\bibfield  {journal} {\bibinfo  {journal} {Nature Photon}\ }\textbf {\bibinfo {volume} {1}},\ \bibinfo {pages} {165} (\bibinfo {year} {2007})}\BibitemShut {NoStop}%
\bibitem [{\citenamefont {Cari{\~n}e}\ \emph {et~al.}(2020)\citenamefont {Cari{\~n}e}, \citenamefont {Ca{\~n}as}, \citenamefont {Skrzypczyk}, \citenamefont {{\v S}upi{\'c}}, \citenamefont {Guerrero}, \citenamefont {Garcia}, \citenamefont {Pereira}, \citenamefont {Prosser}, \citenamefont {Xavier}, \citenamefont {Delgado}, \citenamefont {Walborn}, \citenamefont {Cavalcanti},\ and\ \citenamefont {Lima}}]{Carine_Optica2020}%
  \BibitemOpen
  \bibfield  {author} {\bibinfo {author} {\bibfnamefont {J.}~\bibnamefont {Cari{\~n}e}}, \bibinfo {author} {\bibfnamefont {G.}~\bibnamefont {Ca{\~n}as}}, \bibinfo {author} {\bibfnamefont {P.}~\bibnamefont {Skrzypczyk}}, \bibinfo {author} {\bibfnamefont {I.}~\bibnamefont {{\v S}upi{\'c}}}, \bibinfo {author} {\bibfnamefont {N.}~\bibnamefont {Guerrero}}, \bibinfo {author} {\bibfnamefont {T.}~\bibnamefont {Garcia}}, \bibinfo {author} {\bibfnamefont {L.}~\bibnamefont {Pereira}}, \bibinfo {author} {\bibfnamefont {M.~a.~S.}\ \bibnamefont {Prosser}}, \bibinfo {author} {\bibfnamefont {G.~B.}\ \bibnamefont {Xavier}}, \bibinfo {author} {\bibfnamefont {A.}~\bibnamefont {Delgado}}, \bibinfo {author} {\bibfnamefont {S.~P.}\ \bibnamefont {Walborn}}, \bibinfo {author} {\bibfnamefont {D.}~\bibnamefont {Cavalcanti}},\ and\ \bibinfo {author} {\bibfnamefont {G.}~\bibnamefont {Lima}},\ }\bibfield  {title} {\bibinfo {title} {Multi-core fiber integrated multi-port beam splitters for quantum information processing},\ }\href {https://doi.org/10.1364/OPTICA.388912} {\bibfield  {journal} {\bibinfo  {journal} {Optica, OPTICA}\ }\textbf {\bibinfo {volume} {7}},\ \bibinfo {pages} {542} (\bibinfo {year} {2020})}\BibitemShut {NoStop}%
\bibitem [{\citenamefont {Humphreys}\ \emph {et~al.}(2013)\citenamefont {Humphreys}, \citenamefont {Metcalf}, \citenamefont {Spring}, \citenamefont {Moore}, \citenamefont {Jin}, \citenamefont {Barbieri}, \citenamefont {Kolthammer},\ and\ \citenamefont {Walmsley}}]{Humphreys_Phys.Rev.Lett.2013}%
  \BibitemOpen
  \bibfield  {author} {\bibinfo {author} {\bibfnamefont {P.~C.}\ \bibnamefont {Humphreys}}, \bibinfo {author} {\bibfnamefont {B.~J.}\ \bibnamefont {Metcalf}}, \bibinfo {author} {\bibfnamefont {J.~B.}\ \bibnamefont {Spring}}, \bibinfo {author} {\bibfnamefont {M.}~\bibnamefont {Moore}}, \bibinfo {author} {\bibfnamefont {X.-M.}\ \bibnamefont {Jin}}, \bibinfo {author} {\bibfnamefont {M.}~\bibnamefont {Barbieri}}, \bibinfo {author} {\bibfnamefont {W.~S.}\ \bibnamefont {Kolthammer}},\ and\ \bibinfo {author} {\bibfnamefont {I.~A.}\ \bibnamefont {Walmsley}},\ }\bibfield  {title} {\bibinfo {title} {Linear {{Optical Quantum Computing}} in a {{Single Spatial Mode}}},\ }\href {https://doi.org/10.1103/PhysRevLett.111.150501} {\bibfield  {journal} {\bibinfo  {journal} {Phys. Rev. Lett.}\ }\textbf {\bibinfo {volume} {111}},\ \bibinfo {pages} {150501} (\bibinfo {year} {2013})}\BibitemShut {NoStop}%
\bibitem [{\citenamefont {Lukens}\ and\ \citenamefont {Lougovski}(2017)}]{Lukens_Optica2017}%
  \BibitemOpen
  \bibfield  {author} {\bibinfo {author} {\bibfnamefont {J.~M.}\ \bibnamefont {Lukens}}\ and\ \bibinfo {author} {\bibfnamefont {P.}~\bibnamefont {Lougovski}},\ }\bibfield  {title} {\bibinfo {title} {Frequency-encoded photonic qubits for scalable quantum information processing},\ }\href {https://doi.org/10.1364/OPTICA.4.000008} {\bibfield  {journal} {\bibinfo  {journal} {Optica}\ }\textbf {\bibinfo {volume} {4}},\ \bibinfo {pages} {8} (\bibinfo {year} {2017})}\BibitemShut {NoStop}%
\bibitem [{\citenamefont {Cui}\ \emph {et~al.}(2020)\citenamefont {Cui}, \citenamefont {Seshadreesan}, \citenamefont {Guha},\ and\ \citenamefont {Fan}}]{Cui_Phys.Rev.Lett.2020}%
  \BibitemOpen
  \bibfield  {author} {\bibinfo {author} {\bibfnamefont {C.}~\bibnamefont {Cui}}, \bibinfo {author} {\bibfnamefont {K.~P.}\ \bibnamefont {Seshadreesan}}, \bibinfo {author} {\bibfnamefont {S.}~\bibnamefont {Guha}},\ and\ \bibinfo {author} {\bibfnamefont {L.}~\bibnamefont {Fan}},\ }\bibfield  {title} {\bibinfo {title} {High-{{Dimensional Frequency-Encoded Quantum Information Processing}} with {{Passive Photonics}} and {{Time-Resolving Detection}}},\ }\href {https://doi.org/10.1103/PhysRevLett.124.190502} {\bibfield  {journal} {\bibinfo  {journal} {Phys. Rev. Lett.}\ }\textbf {\bibinfo {volume} {124}},\ \bibinfo {pages} {190502} (\bibinfo {year} {2020})}\BibitemShut {NoStop}%
\bibitem [{\citenamefont {Brecht}\ \emph {et~al.}(2015)\citenamefont {Brecht}, \citenamefont {Reddy}, \citenamefont {Silberhorn},\ and\ \citenamefont {Raymer}}]{Brecht_Phys.Rev.X2015}%
  \BibitemOpen
  \bibfield  {author} {\bibinfo {author} {\bibfnamefont {B.}~\bibnamefont {Brecht}}, \bibinfo {author} {\bibfnamefont {D.~V.}\ \bibnamefont {Reddy}}, \bibinfo {author} {\bibfnamefont {C.}~\bibnamefont {Silberhorn}},\ and\ \bibinfo {author} {\bibfnamefont {M.~G.}\ \bibnamefont {Raymer}},\ }\bibfield  {title} {\bibinfo {title} {Photon {{Temporal Modes}}: {{A Complete Framework}} for {{Quantum Information Science}}},\ }\href {https://doi.org/10.1103/PhysRevX.5.041017} {\bibfield  {journal} {\bibinfo  {journal} {Phys. Rev. X}\ }\textbf {\bibinfo {volume} {5}},\ \bibinfo {pages} {041017} (\bibinfo {year} {2015})}\BibitemShut {NoStop}%
\bibitem [{\citenamefont {Fabre}\ \emph {et~al.}(2022)\citenamefont {Fabre}, \citenamefont {Keller},\ and\ \citenamefont {Milman}}]{Fabre_Phys.Rev.A2022}%
  \BibitemOpen
  \bibfield  {author} {\bibinfo {author} {\bibfnamefont {N.}~\bibnamefont {Fabre}}, \bibinfo {author} {\bibfnamefont {A.}~\bibnamefont {Keller}},\ and\ \bibinfo {author} {\bibfnamefont {P.}~\bibnamefont {Milman}},\ }\bibfield  {title} {\bibinfo {title} {Time and frequency as quantum continuous variables},\ }\href {https://doi.org/10.1103/PhysRevA.105.052429} {\bibfield  {journal} {\bibinfo  {journal} {Phys. Rev. A}\ }\textbf {\bibinfo {volume} {105}},\ \bibinfo {pages} {052429} (\bibinfo {year} {2022})}\BibitemShut {NoStop}%
\bibitem [{\citenamefont {Fabre}\ \emph {et~al.}(2020)\citenamefont {Fabre}, \citenamefont {Maltese}, \citenamefont {Appas}, \citenamefont {Felicetti}, \citenamefont {Ketterer}, \citenamefont {Keller}, \citenamefont {Coudreau}, \citenamefont {Baboux}, \citenamefont {Amanti}, \citenamefont {Ducci},\ and\ \citenamefont {Milman}}]{Fabre_Phys.Rev.A2020a}%
  \BibitemOpen
  \bibfield  {author} {\bibinfo {author} {\bibfnamefont {N.}~\bibnamefont {Fabre}}, \bibinfo {author} {\bibfnamefont {G.}~\bibnamefont {Maltese}}, \bibinfo {author} {\bibfnamefont {F.}~\bibnamefont {Appas}}, \bibinfo {author} {\bibfnamefont {S.}~\bibnamefont {Felicetti}}, \bibinfo {author} {\bibfnamefont {A.}~\bibnamefont {Ketterer}}, \bibinfo {author} {\bibfnamefont {A.}~\bibnamefont {Keller}}, \bibinfo {author} {\bibfnamefont {T.}~\bibnamefont {Coudreau}}, \bibinfo {author} {\bibfnamefont {F.}~\bibnamefont {Baboux}}, \bibinfo {author} {\bibfnamefont {M.~I.}\ \bibnamefont {Amanti}}, \bibinfo {author} {\bibfnamefont {S.}~\bibnamefont {Ducci}},\ and\ \bibinfo {author} {\bibfnamefont {P.}~\bibnamefont {Milman}},\ }\bibfield  {title} {\bibinfo {title} {Generation of a time-frequency grid state with integrated biphoton frequency combs},\ }\href {https://doi.org/10.1103/PhysRevA.102.012607} {\bibfield  {journal} {\bibinfo  {journal} {Phys. Rev. A}\ }\textbf {\bibinfo {volume} {102}},\ \bibinfo {pages} {012607} (\bibinfo {year} {2020})}\BibitemShut {NoStop}%
\bibitem [{\citenamefont {Yamazaki}\ \emph {et~al.}(2023{\natexlab{a}})\citenamefont {Yamazaki}, \citenamefont {Arizono}, \citenamefont {Kobayashi}, \citenamefont {Ikuta},\ and\ \citenamefont {Yamamoto}}]{Yamazaki_Phys.Rev.Lett.2023}%
  \BibitemOpen
  \bibfield  {author} {\bibinfo {author} {\bibfnamefont {T.}~\bibnamefont {Yamazaki}}, \bibinfo {author} {\bibfnamefont {T.}~\bibnamefont {Arizono}}, \bibinfo {author} {\bibfnamefont {T.}~\bibnamefont {Kobayashi}}, \bibinfo {author} {\bibfnamefont {R.}~\bibnamefont {Ikuta}},\ and\ \bibinfo {author} {\bibfnamefont {T.}~\bibnamefont {Yamamoto}},\ }\bibfield  {title} {\bibinfo {title} {Linear {{Optical Quantum Computation}} with {{Frequency-Comb Qubits}} and {{Passive Devices}}},\ }\href {https://doi.org/10.1103/PhysRevLett.130.200602} {\bibfield  {journal} {\bibinfo  {journal} {Phys. Rev. Lett.}\ }\textbf {\bibinfo {volume} {130}},\ \bibinfo {pages} {200602} (\bibinfo {year} {2023}{\natexlab{a}})}\BibitemShut {NoStop}%
\bibitem [{\citenamefont {Vaziri}\ \emph {et~al.}(2002)\citenamefont {Vaziri}, \citenamefont {Weihs},\ and\ \citenamefont {Zeilinger}}]{Vaziri_Phys.Rev.Lett.2002}%
  \BibitemOpen
  \bibfield  {author} {\bibinfo {author} {\bibfnamefont {A.}~\bibnamefont {Vaziri}}, \bibinfo {author} {\bibfnamefont {G.}~\bibnamefont {Weihs}},\ and\ \bibinfo {author} {\bibfnamefont {A.}~\bibnamefont {Zeilinger}},\ }\bibfield  {title} {\bibinfo {title} {Experimental {{Two-Photon}}, {{Three-Dimensional Entanglement}} for {{Quantum Communication}}},\ }\href {https://doi.org/10.1103/PhysRevLett.89.240401} {\bibfield  {journal} {\bibinfo  {journal} {Phys. Rev. Lett.}\ }\textbf {\bibinfo {volume} {89}},\ \bibinfo {pages} {240401} (\bibinfo {year} {2002})}\BibitemShut {NoStop}%
\bibitem [{\citenamefont {Erhard}\ \emph {et~al.}(2018)\citenamefont {Erhard}, \citenamefont {Fickler}, \citenamefont {Krenn},\ and\ \citenamefont {Zeilinger}}]{Erhard_LightSci.Appl.2018}%
  \BibitemOpen
  \bibfield  {author} {\bibinfo {author} {\bibfnamefont {M.}~\bibnamefont {Erhard}}, \bibinfo {author} {\bibfnamefont {R.}~\bibnamefont {Fickler}}, \bibinfo {author} {\bibfnamefont {M.}~\bibnamefont {Krenn}},\ and\ \bibinfo {author} {\bibfnamefont {A.}~\bibnamefont {Zeilinger}},\ }\bibfield  {title} {\bibinfo {title} {Twisted photons: New quantum perspectives in high dimensions},\ }\href {https://doi.org/10.1038/lsa.2017.146} {\bibfield  {journal} {\bibinfo  {journal} {Light Sci Appl}\ }\textbf {\bibinfo {volume} {7}},\ \bibinfo {pages} {17146} (\bibinfo {year} {2018})}\BibitemShut {NoStop}%
\bibitem [{\citenamefont {Collins}\ \emph {et~al.}(2002)\citenamefont {Collins}, \citenamefont {Gisin}, \citenamefont {Linden}, \citenamefont {Massar},\ and\ \citenamefont {Popescu}}]{Collins_Phys.Rev.Lett.2002}%
  \BibitemOpen
  \bibfield  {author} {\bibinfo {author} {\bibfnamefont {D.}~\bibnamefont {Collins}}, \bibinfo {author} {\bibfnamefont {N.}~\bibnamefont {Gisin}}, \bibinfo {author} {\bibfnamefont {N.}~\bibnamefont {Linden}}, \bibinfo {author} {\bibfnamefont {S.}~\bibnamefont {Massar}},\ and\ \bibinfo {author} {\bibfnamefont {S.}~\bibnamefont {Popescu}},\ }\bibfield  {title} {\bibinfo {title} {Bell {{Inequalities}} for {{Arbitrarily High-Dimensional Systems}}},\ }\href {https://doi.org/10.1103/PhysRevLett.88.040404} {\bibfield  {journal} {\bibinfo  {journal} {Phys. Rev. Lett.}\ }\textbf {\bibinfo {volume} {88}},\ \bibinfo {pages} {040404} (\bibinfo {year} {2002})}\BibitemShut {NoStop}%
\bibitem [{\citenamefont {Cerf}\ \emph {et~al.}(2002)\citenamefont {Cerf}, \citenamefont {Bourennane}, \citenamefont {Karlsson},\ and\ \citenamefont {Gisin}}]{Cerf_Phys.Rev.Lett.2002}%
  \BibitemOpen
  \bibfield  {author} {\bibinfo {author} {\bibfnamefont {N.~J.}\ \bibnamefont {Cerf}}, \bibinfo {author} {\bibfnamefont {M.}~\bibnamefont {Bourennane}}, \bibinfo {author} {\bibfnamefont {A.}~\bibnamefont {Karlsson}},\ and\ \bibinfo {author} {\bibfnamefont {N.}~\bibnamefont {Gisin}},\ }\bibfield  {title} {\bibinfo {title} {Security of quantum key distribution using d-level systems},\ }\href {https://doi.org/10.1103/PhysRevLett.88.127902} {\bibfield  {journal} {\bibinfo  {journal} {Phys. Rev. Lett.}\ }\textbf {\bibinfo {volume} {88}},\ \bibinfo {pages} {127902} (\bibinfo {year} {2002})}\BibitemShut {NoStop}%
\bibitem [{\citenamefont {Zhu}\ \emph {et~al.}(2021)\citenamefont {Zhu}, \citenamefont {Tyler}, \citenamefont {Valencia}, \citenamefont {Malik},\ and\ \citenamefont {Leach}}]{Zhu_AVSQuantumSci.2021}%
  \BibitemOpen
  \bibfield  {author} {\bibinfo {author} {\bibfnamefont {F.}~\bibnamefont {Zhu}}, \bibinfo {author} {\bibfnamefont {M.}~\bibnamefont {Tyler}}, \bibinfo {author} {\bibfnamefont {N.~H.}\ \bibnamefont {Valencia}}, \bibinfo {author} {\bibfnamefont {M.}~\bibnamefont {Malik}},\ and\ \bibinfo {author} {\bibfnamefont {J.}~\bibnamefont {Leach}},\ }\bibfield  {title} {\bibinfo {title} {Is high-dimensional photonic entanglement robust to noise?},\ }\href {https://doi.org/10.1116/5.0033889} {\bibfield  {journal} {\bibinfo  {journal} {AVS Quantum Sci.}\ }\textbf {\bibinfo {volume} {3}},\ \bibinfo {pages} {011401} (\bibinfo {year} {2021})}\BibitemShut {NoStop}%
\bibitem [{\citenamefont {Cozzolino}\ \emph {et~al.}(2019)\citenamefont {Cozzolino}, \citenamefont {Da~Lio}, \citenamefont {Bacco},\ and\ \citenamefont {Oxenl{\o}we}}]{Cozzolino_Adv.QuantumTechnol.2019}%
  \BibitemOpen
  \bibfield  {author} {\bibinfo {author} {\bibfnamefont {D.}~\bibnamefont {Cozzolino}}, \bibinfo {author} {\bibfnamefont {B.}~\bibnamefont {Da~Lio}}, \bibinfo {author} {\bibfnamefont {D.}~\bibnamefont {Bacco}},\ and\ \bibinfo {author} {\bibfnamefont {L.~K.}\ \bibnamefont {Oxenl{\o}we}},\ }\bibfield  {title} {\bibinfo {title} {High-{{Dimensional Quantum Communication}}: {{Benefits}}, {{Progress}}, and {{Future Challenges}}},\ }\href {https://doi.org/10.1002/qute.201900038} {\bibfield  {journal} {\bibinfo  {journal} {Advanced Quantum Technologies}\ }\textbf {\bibinfo {volume} {2}},\ \bibinfo {pages} {1900038} (\bibinfo {year} {2019})}\BibitemShut {NoStop}%
\bibitem [{\citenamefont {Browne}\ and\ \citenamefont {Rudolph}(2005)}]{Browne_Phys.Rev.Lett.2005}%
  \BibitemOpen
  \bibfield  {author} {\bibinfo {author} {\bibfnamefont {D.~E.}\ \bibnamefont {Browne}}\ and\ \bibinfo {author} {\bibfnamefont {T.}~\bibnamefont {Rudolph}},\ }\bibfield  {title} {\bibinfo {title} {Resource-{{Efficient Linear Optical Quantum Computation}}},\ }\href {https://doi.org/10.1103/PhysRevLett.95.010501} {\bibfield  {journal} {\bibinfo  {journal} {Phys. Rev. Lett.}\ }\textbf {\bibinfo {volume} {95}},\ \bibinfo {pages} {010501} (\bibinfo {year} {2005})}\BibitemShut {NoStop}%
\bibitem [{\citenamefont {Kieling}\ \emph {et~al.}(2007)\citenamefont {Kieling}, \citenamefont {Rudolph},\ and\ \citenamefont {Eisert}}]{Kieling_Phys.Rev.Lett.2007}%
  \BibitemOpen
  \bibfield  {author} {\bibinfo {author} {\bibfnamefont {K.}~\bibnamefont {Kieling}}, \bibinfo {author} {\bibfnamefont {T.}~\bibnamefont {Rudolph}},\ and\ \bibinfo {author} {\bibfnamefont {J.}~\bibnamefont {Eisert}},\ }\bibfield  {title} {\bibinfo {title} {Percolation, {{Renormalization}}, and {{Quantum Computing}} with {{Nondeterministic Gates}}},\ }\href {https://doi.org/10.1103/PhysRevLett.99.130501} {\bibfield  {journal} {\bibinfo  {journal} {Phys. Rev. Lett.}\ }\textbf {\bibinfo {volume} {99}},\ \bibinfo {pages} {130501} (\bibinfo {year} {2007})}\BibitemShut {NoStop}%
\bibitem [{\citenamefont {{Gimeno-Segovia}}\ \emph {et~al.}(2015)\citenamefont {{Gimeno-Segovia}}, \citenamefont {Shadbolt}, \citenamefont {Browne},\ and\ \citenamefont {Rudolph}}]{Gimeno-Segovia_Phys.Rev.Lett.2015}%
  \BibitemOpen
  \bibfield  {author} {\bibinfo {author} {\bibfnamefont {M.}~\bibnamefont {{Gimeno-Segovia}}}, \bibinfo {author} {\bibfnamefont {P.}~\bibnamefont {Shadbolt}}, \bibinfo {author} {\bibfnamefont {D.~E.}\ \bibnamefont {Browne}},\ and\ \bibinfo {author} {\bibfnamefont {T.}~\bibnamefont {Rudolph}},\ }\bibfield  {title} {\bibinfo {title} {From {{Three-Photon Greenberger-Horne-Zeilinger States}} to {{Ballistic Universal Quantum Computation}}},\ }\href {https://doi.org/10.1103/PhysRevLett.115.020502} {\bibfield  {journal} {\bibinfo  {journal} {Phys. Rev. Lett.}\ }\textbf {\bibinfo {volume} {115}},\ \bibinfo {pages} {020502} (\bibinfo {year} {2015})}\BibitemShut {NoStop}%
\bibitem [{\citenamefont {Bartolucci}\ \emph {et~al.}(2023)\citenamefont {Bartolucci}, \citenamefont {Birchall}, \citenamefont {Bomb{\'i}n}, \citenamefont {Cable}, \citenamefont {Dawson}, \citenamefont {{Gimeno-Segovia}}, \citenamefont {Johnston}, \citenamefont {Kieling}, \citenamefont {Nickerson}, \citenamefont {Pant}, \citenamefont {Pastawski}, \citenamefont {Rudolph},\ and\ \citenamefont {Sparrow}}]{Bartolucci_Nat.Commun.2023}%
  \BibitemOpen
  \bibfield  {author} {\bibinfo {author} {\bibfnamefont {S.}~\bibnamefont {Bartolucci}}, \bibinfo {author} {\bibfnamefont {P.}~\bibnamefont {Birchall}}, \bibinfo {author} {\bibfnamefont {H.}~\bibnamefont {Bomb{\'i}n}}, \bibinfo {author} {\bibfnamefont {H.}~\bibnamefont {Cable}}, \bibinfo {author} {\bibfnamefont {C.}~\bibnamefont {Dawson}}, \bibinfo {author} {\bibfnamefont {M.}~\bibnamefont {{Gimeno-Segovia}}}, \bibinfo {author} {\bibfnamefont {E.}~\bibnamefont {Johnston}}, \bibinfo {author} {\bibfnamefont {K.}~\bibnamefont {Kieling}}, \bibinfo {author} {\bibfnamefont {N.}~\bibnamefont {Nickerson}}, \bibinfo {author} {\bibfnamefont {M.}~\bibnamefont {Pant}}, \bibinfo {author} {\bibfnamefont {F.}~\bibnamefont {Pastawski}}, \bibinfo {author} {\bibfnamefont {T.}~\bibnamefont {Rudolph}},\ and\ \bibinfo {author} {\bibfnamefont {C.}~\bibnamefont {Sparrow}},\ }\bibfield  {title} {\bibinfo {title} {Fusion-based quantum computation},\ }\href {https://doi.org/10.1038/s41467-023-36493-1} {\bibfield  {journal} {\bibinfo  {journal} {Nat Commun}\ }\textbf {\bibinfo {volume} {14}},\ \bibinfo {pages} {912} (\bibinfo {year} {2023})}\BibitemShut {NoStop}%
\bibitem [{\citenamefont {Sahay}\ \emph {et~al.}(2023)\citenamefont {Sahay}, \citenamefont {Claes},\ and\ \citenamefont {Puri}}]{Sahay_Phys.Rev.Lett.2023}%
  \BibitemOpen
  \bibfield  {author} {\bibinfo {author} {\bibfnamefont {K.}~\bibnamefont {Sahay}}, \bibinfo {author} {\bibfnamefont {J.}~\bibnamefont {Claes}},\ and\ \bibinfo {author} {\bibfnamefont {S.}~\bibnamefont {Puri}},\ }\bibfield  {title} {\bibinfo {title} {Tailoring {{Fusion-Based Error Correction}} for {{High Thresholds}} to {{Biased Fusion Failures}}},\ }\href {https://doi.org/10.1103/PhysRevLett.131.120604} {\bibfield  {journal} {\bibinfo  {journal} {Phys. Rev. Lett.}\ }\textbf {\bibinfo {volume} {131}},\ \bibinfo {pages} {120604} (\bibinfo {year} {2023})}\BibitemShut {NoStop}%
\bibitem [{\citenamefont {Azuma}\ \emph {et~al.}(2015)\citenamefont {Azuma}, \citenamefont {Tamaki},\ and\ \citenamefont {Lo}}]{Azuma_Nat.Commun.2015}%
  \BibitemOpen
  \bibfield  {author} {\bibinfo {author} {\bibfnamefont {K.}~\bibnamefont {Azuma}}, \bibinfo {author} {\bibfnamefont {K.}~\bibnamefont {Tamaki}},\ and\ \bibinfo {author} {\bibfnamefont {H.-K.}\ \bibnamefont {Lo}},\ }\bibfield  {title} {\bibinfo {title} {All-photonic quantum repeaters},\ }\href {https://doi.org/10.1038/ncomms7787} {\bibfield  {journal} {\bibinfo  {journal} {Nat Commun}\ }\textbf {\bibinfo {volume} {6}},\ \bibinfo {pages} {6787} (\bibinfo {year} {2015})}\BibitemShut {NoStop}%
\bibitem [{\citenamefont {Ewert}\ \emph {et~al.}(2016)\citenamefont {Ewert}, \citenamefont {Bergmann},\ and\ \citenamefont {{van Loock}}}]{Ewert_Phys.Rev.Lett.2016}%
  \BibitemOpen
  \bibfield  {author} {\bibinfo {author} {\bibfnamefont {F.}~\bibnamefont {Ewert}}, \bibinfo {author} {\bibfnamefont {M.}~\bibnamefont {Bergmann}},\ and\ \bibinfo {author} {\bibfnamefont {P.}~\bibnamefont {{van Loock}}},\ }\bibfield  {title} {\bibinfo {title} {Ultrafast {{Long-Distance Quantum Communication}} with {{Static Linear Optics}}},\ }\href {https://doi.org/10.1103/PhysRevLett.117.210501} {\bibfield  {journal} {\bibinfo  {journal} {Phys. Rev. Lett.}\ }\textbf {\bibinfo {volume} {117}},\ \bibinfo {pages} {210501} (\bibinfo {year} {2016})},\ \Eprint {https://arxiv.org/abs/1503.06777} {arXiv:1503.06777} \BibitemShut {NoStop}%
\bibitem [{\citenamefont {Weinfurter}(1994)}]{Weinfurter_EPL1994}%
  \BibitemOpen
  \bibfield  {author} {\bibinfo {author} {\bibfnamefont {H.}~\bibnamefont {Weinfurter}},\ }\bibfield  {title} {\bibinfo {title} {Experimental bell-state analysis},\ }\href {https://doi.org/10.1209/0295-5075/25/8/001} {\bibfield  {journal} {\bibinfo  {journal} {EPL}\ }\textbf {\bibinfo {volume} {25}},\ \bibinfo {pages} {559} (\bibinfo {year} {1994})}\BibitemShut {NoStop}%
\bibitem [{\citenamefont {Braunstein}\ and\ \citenamefont {Mann}(1995)}]{Braunstein_Phys.Rev.A1995}%
  \BibitemOpen
  \bibfield  {author} {\bibinfo {author} {\bibfnamefont {S.~L.}\ \bibnamefont {Braunstein}}\ and\ \bibinfo {author} {\bibfnamefont {A.}~\bibnamefont {Mann}},\ }\bibfield  {title} {\bibinfo {title} {Measurement of the {{Bell}} operator and quantum teleportation},\ }\href {https://doi.org/10.1103/PhysRevA.51.R1727} {\bibfield  {journal} {\bibinfo  {journal} {Phys. Rev. A}\ }\textbf {\bibinfo {volume} {51}},\ \bibinfo {pages} {R1727} (\bibinfo {year} {1995})}\BibitemShut {NoStop}%
\bibitem [{\citenamefont {Calsamiglia}\ and\ \citenamefont {L{\"u}tkenhaus}(2001)}]{Calsamiglia_Appl.Phys.B2001}%
  \BibitemOpen
  \bibfield  {author} {\bibinfo {author} {\bibfnamefont {J.}~\bibnamefont {Calsamiglia}}\ and\ \bibinfo {author} {\bibfnamefont {N.}~\bibnamefont {L{\"u}tkenhaus}},\ }\bibfield  {title} {\bibinfo {title} {Maximum efficiency of a linear-optical {{Bell-state}} analyzer},\ }\href {https://doi.org/10.1007/s003400000484} {\bibfield  {journal} {\bibinfo  {journal} {Appl Phys B}\ }\textbf {\bibinfo {volume} {72}},\ \bibinfo {pages} {67} (\bibinfo {year} {2001})}\BibitemShut {NoStop}%
\bibitem [{\citenamefont {Calsamiglia}(2002)}]{Calsamiglia_Phys.Rev.A2002}%
  \BibitemOpen
  \bibfield  {author} {\bibinfo {author} {\bibfnamefont {J.}~\bibnamefont {Calsamiglia}},\ }\bibfield  {title} {\bibinfo {title} {Generalized measurements by linear elements},\ }\href {https://doi.org/10.1103/PhysRevA.65.030301} {\bibfield  {journal} {\bibinfo  {journal} {Phys. Rev. A}\ }\textbf {\bibinfo {volume} {65}},\ \bibinfo {pages} {030301} (\bibinfo {year} {2002})}\BibitemShut {NoStop}%
\bibitem [{\citenamefont {Luo}\ \emph {et~al.}(2019)\citenamefont {Luo}, \citenamefont {Zhong}, \citenamefont {Erhard}, \citenamefont {Wang}, \citenamefont {Peng}, \citenamefont {Krenn}, \citenamefont {Jiang}, \citenamefont {Li}, \citenamefont {Liu}, \citenamefont {Lu}, \citenamefont {Zeilinger},\ and\ \citenamefont {Pan}}]{Luo_Phys.Rev.Lett.2019}%
  \BibitemOpen
  \bibfield  {author} {\bibinfo {author} {\bibfnamefont {Y.-H.}\ \bibnamefont {Luo}}, \bibinfo {author} {\bibfnamefont {H.-S.}\ \bibnamefont {Zhong}}, \bibinfo {author} {\bibfnamefont {M.}~\bibnamefont {Erhard}}, \bibinfo {author} {\bibfnamefont {X.-L.}\ \bibnamefont {Wang}}, \bibinfo {author} {\bibfnamefont {L.-C.}\ \bibnamefont {Peng}}, \bibinfo {author} {\bibfnamefont {M.}~\bibnamefont {Krenn}}, \bibinfo {author} {\bibfnamefont {X.}~\bibnamefont {Jiang}}, \bibinfo {author} {\bibfnamefont {L.}~\bibnamefont {Li}}, \bibinfo {author} {\bibfnamefont {N.-L.}\ \bibnamefont {Liu}}, \bibinfo {author} {\bibfnamefont {C.-Y.}\ \bibnamefont {Lu}}, \bibinfo {author} {\bibfnamefont {A.}~\bibnamefont {Zeilinger}},\ and\ \bibinfo {author} {\bibfnamefont {J.-W.}\ \bibnamefont {Pan}},\ }\bibfield  {title} {\bibinfo {title} {Quantum {{Teleportation}} in {{High Dimensions}}},\ }\href {https://doi.org/10.1103/PhysRevLett.123.070505} {\bibfield  {journal} {\bibinfo  {journal} {Phys. Rev. Lett.}\ }\textbf {\bibinfo {volume} {123}},\ \bibinfo {pages} {070505} (\bibinfo {year} {2019})}\BibitemShut {NoStop}%
\bibitem [{\citenamefont {Zhang}\ \emph {et~al.}(2019)\citenamefont {Zhang}, \citenamefont {Zhang}, \citenamefont {Hu}, \citenamefont {Liu}, \citenamefont {Huang}, \citenamefont {Li},\ and\ \citenamefont {Guo}}]{Zhang_Phys.Rev.A2019a}%
  \BibitemOpen
  \bibfield  {author} {\bibinfo {author} {\bibfnamefont {H.}~\bibnamefont {Zhang}}, \bibinfo {author} {\bibfnamefont {C.}~\bibnamefont {Zhang}}, \bibinfo {author} {\bibfnamefont {X.-M.}\ \bibnamefont {Hu}}, \bibinfo {author} {\bibfnamefont {B.-H.}\ \bibnamefont {Liu}}, \bibinfo {author} {\bibfnamefont {Y.-F.}\ \bibnamefont {Huang}}, \bibinfo {author} {\bibfnamefont {C.-F.}\ \bibnamefont {Li}},\ and\ \bibinfo {author} {\bibfnamefont {G.-C.}\ \bibnamefont {Guo}},\ }\bibfield  {title} {\bibinfo {title} {Arbitrary two-particle high-dimensional {{Bell-state}} measurement by auxiliary entanglement},\ }\href {https://doi.org/10.1103/PhysRevA.99.052301} {\bibfield  {journal} {\bibinfo  {journal} {Phys. Rev. A}\ }\textbf {\bibinfo {volume} {99}},\ \bibinfo {pages} {052301} (\bibinfo {year} {2019})}\BibitemShut {NoStop}%
\bibitem [{\citenamefont {Bharos}\ \emph {et~al.}(2024)\citenamefont {Bharos}, \citenamefont {Markovich},\ and\ \citenamefont {Borregaard}}]{Bharos_2024}%
  \BibitemOpen
  \bibfield  {author} {\bibinfo {author} {\bibfnamefont {N.}~\bibnamefont {Bharos}}, \bibinfo {author} {\bibfnamefont {L.}~\bibnamefont {Markovich}},\ and\ \bibinfo {author} {\bibfnamefont {J.}~\bibnamefont {Borregaard}},\ }\href@noop {} {\bibinfo {title} {Efficient {{High-Dimensional Entangled State Analyzer}} with {{Linear Optics}}}} (\bibinfo {year} {2024}),\ \Eprint {https://arxiv.org/abs/2401.15066} {arXiv:2401.15066} \BibitemShut {NoStop}%
\bibitem [{\citenamefont {Paesani}\ \emph {et~al.}(2021)\citenamefont {Paesani}, \citenamefont {Bulmer}, \citenamefont {Jones}, \citenamefont {Santagati},\ and\ \citenamefont {Laing}}]{Paesani_Phys.Rev.Lett.2021}%
  \BibitemOpen
  \bibfield  {author} {\bibinfo {author} {\bibfnamefont {S.}~\bibnamefont {Paesani}}, \bibinfo {author} {\bibfnamefont {J.~F.~F.}\ \bibnamefont {Bulmer}}, \bibinfo {author} {\bibfnamefont {A.~E.}\ \bibnamefont {Jones}}, \bibinfo {author} {\bibfnamefont {R.}~\bibnamefont {Santagati}},\ and\ \bibinfo {author} {\bibfnamefont {A.}~\bibnamefont {Laing}},\ }\bibfield  {title} {\bibinfo {title} {Scheme for {{Universal High-Dimensional Quantum Computation}} with {{Linear Optics}}},\ }\href {https://doi.org/10.1103/PhysRevLett.126.230504} {\bibfield  {journal} {\bibinfo  {journal} {Phys. Rev. Lett.}\ }\textbf {\bibinfo {volume} {126}},\ \bibinfo {pages} {230504} (\bibinfo {year} {2021})}\BibitemShut {NoStop}%
\bibitem [{\citenamefont {Grice}(2011)}]{Grice_Phys.Rev.A2011}%
  \BibitemOpen
  \bibfield  {author} {\bibinfo {author} {\bibfnamefont {W.~P.}\ \bibnamefont {Grice}},\ }\bibfield  {title} {\bibinfo {title} {Arbitrarily complete {{Bell-state}} measurement using only linear optical elements},\ }\href {https://doi.org/10.1103/PhysRevA.84.042331} {\bibfield  {journal} {\bibinfo  {journal} {Phys. Rev. A}\ }\textbf {\bibinfo {volume} {84}},\ \bibinfo {pages} {042331} (\bibinfo {year} {2011})}\BibitemShut {NoStop}%
\bibitem [{\citenamefont {Zaidi}\ and\ \citenamefont {Van~Loock}(2013)}]{Zaidi_Phys.Rev.Lett.2013}%
  \BibitemOpen
  \bibfield  {author} {\bibinfo {author} {\bibfnamefont {H.~A.}\ \bibnamefont {Zaidi}}\ and\ \bibinfo {author} {\bibfnamefont {P.}~\bibnamefont {Van~Loock}},\ }\bibfield  {title} {\bibinfo {title} {Beating the {{One-Half Limit}} of {{Ancilla-Free Linear Optics Bell Measurements}}},\ }\href {https://doi.org/10.1103/PhysRevLett.110.260501} {\bibfield  {journal} {\bibinfo  {journal} {Phys. Rev. Lett.}\ }\textbf {\bibinfo {volume} {110}},\ \bibinfo {pages} {260501} (\bibinfo {year} {2013})}\BibitemShut {NoStop}%
\bibitem [{\citenamefont {Ewert}\ and\ \citenamefont {{van Loock}}(2014)}]{Ewert_Phys.Rev.Lett.2014}%
  \BibitemOpen
  \bibfield  {author} {\bibinfo {author} {\bibfnamefont {F.}~\bibnamefont {Ewert}}\ and\ \bibinfo {author} {\bibfnamefont {P.}~\bibnamefont {{van Loock}}},\ }\bibfield  {title} {\bibinfo {title} {3/4-{{Efficient Bell}} measurement with passive linear optics and unentangled ancillae},\ }\href {https://doi.org/10.1103/PhysRevLett.113.140403} {\bibfield  {journal} {\bibinfo  {journal} {Phys. Rev. Lett.}\ }\textbf {\bibinfo {volume} {113}},\ \bibinfo {pages} {140403} (\bibinfo {year} {2014})}\BibitemShut {NoStop}%
\bibitem [{\citenamefont {Olivo}\ and\ \citenamefont {Grosshans}(2018)}]{Olivo_Phys.Rev.A2018}%
  \BibitemOpen
  \bibfield  {author} {\bibinfo {author} {\bibfnamefont {A.}~\bibnamefont {Olivo}}\ and\ \bibinfo {author} {\bibfnamefont {F.}~\bibnamefont {Grosshans}},\ }\bibfield  {title} {\bibinfo {title} {Ancilla-assisted linear optical {{Bell}} measurements and their optimality},\ }\href {https://doi.org/10.1103/PhysRevA.98.042323} {\bibfield  {journal} {\bibinfo  {journal} {Phys. Rev. A}\ }\textbf {\bibinfo {volume} {98}},\ \bibinfo {pages} {042323} (\bibinfo {year} {2018})}\BibitemShut {NoStop}%
\bibitem [{\citenamefont {Bartolucci}\ \emph {et~al.}(2021)\citenamefont {Bartolucci}, \citenamefont {Birchall}, \citenamefont {{Gimeno-Segovia}}, \citenamefont {Johnston}, \citenamefont {Kieling}, \citenamefont {Pant}, \citenamefont {Rudolph}, \citenamefont {Smith}, \citenamefont {Sparrow},\ and\ \citenamefont {Vidrighin}}]{Bartolucci_2021b}%
  \BibitemOpen
  \bibfield  {author} {\bibinfo {author} {\bibfnamefont {S.}~\bibnamefont {Bartolucci}}, \bibinfo {author} {\bibfnamefont {P.~M.}\ \bibnamefont {Birchall}}, \bibinfo {author} {\bibfnamefont {M.}~\bibnamefont {{Gimeno-Segovia}}}, \bibinfo {author} {\bibfnamefont {E.}~\bibnamefont {Johnston}}, \bibinfo {author} {\bibfnamefont {K.}~\bibnamefont {Kieling}}, \bibinfo {author} {\bibfnamefont {M.}~\bibnamefont {Pant}}, \bibinfo {author} {\bibfnamefont {T.}~\bibnamefont {Rudolph}}, \bibinfo {author} {\bibfnamefont {J.}~\bibnamefont {Smith}}, \bibinfo {author} {\bibfnamefont {C.}~\bibnamefont {Sparrow}},\ and\ \bibinfo {author} {\bibfnamefont {M.~D.}\ \bibnamefont {Vidrighin}},\ }\href {https://doi.org/10.48550/arXiv.2106.13825} {\bibinfo {title} {Creation of {{Entangled Photonic States Using Linear Optics}}}} (\bibinfo {year} {2021}),\ \Eprint {https://arxiv.org/abs/2106.13825} {arXiv:2106.13825} \BibitemShut {NoStop}%
\bibitem [{Note1()}]{Note1}%
  \BibitemOpen
  \bibinfo {note} {A ququart is a qudit of dimension $d=4$. The original scheme, called hyperentanglement-assisted Bell-state analysis, is proposed in terms of hyperentanglement, where each photon forms two qubits with two degrees of freedom of the photon.}\BibitemShut {Stop}%
\bibitem [{\citenamefont {Walborn}\ \emph {et~al.}(2003)\citenamefont {Walborn}, \citenamefont {P{\'a}dua},\ and\ \citenamefont {Monken}}]{Walborn_Phys.Rev.A2003}%
  \BibitemOpen
  \bibfield  {author} {\bibinfo {author} {\bibfnamefont {S.~P.}\ \bibnamefont {Walborn}}, \bibinfo {author} {\bibfnamefont {S.}~\bibnamefont {P{\'a}dua}},\ and\ \bibinfo {author} {\bibfnamefont {C.~H.}\ \bibnamefont {Monken}},\ }\bibfield  {title} {\bibinfo {title} {Hyperentanglement-assisted {{Bell-state}} analysis},\ }\href {https://doi.org/10.1103/PhysRevA.68.042313} {\bibfield  {journal} {\bibinfo  {journal} {Phys. Rev. A}\ }\textbf {\bibinfo {volume} {68}},\ \bibinfo {pages} {042313} (\bibinfo {year} {2003})}\BibitemShut {NoStop}%
\bibitem [{\citenamefont {Schuck}\ \emph {et~al.}(2006)\citenamefont {Schuck}, \citenamefont {Huber}, \citenamefont {Kurtsiefer},\ and\ \citenamefont {Weinfurter}}]{Schuck_Phys.Rev.Lett.2006}%
  \BibitemOpen
  \bibfield  {author} {\bibinfo {author} {\bibfnamefont {C.}~\bibnamefont {Schuck}}, \bibinfo {author} {\bibfnamefont {G.}~\bibnamefont {Huber}}, \bibinfo {author} {\bibfnamefont {C.}~\bibnamefont {Kurtsiefer}},\ and\ \bibinfo {author} {\bibfnamefont {H.}~\bibnamefont {Weinfurter}},\ }\bibfield  {title} {\bibinfo {title} {Complete {{Deterministic Linear Optics Bell State Analysis}}},\ }\href {https://doi.org/10.1103/PhysRevLett.96.190501} {\bibfield  {journal} {\bibinfo  {journal} {Phys. Rev. Lett.}\ }\textbf {\bibinfo {volume} {96}},\ \bibinfo {pages} {190501} (\bibinfo {year} {2006})}\BibitemShut {NoStop}%
\bibitem [{\citenamefont {Barreiro}\ \emph {et~al.}(2008)\citenamefont {Barreiro}, \citenamefont {Wei},\ and\ \citenamefont {Kwiat}}]{Barreiro_Nat.Phys.2008}%
  \BibitemOpen
  \bibfield  {author} {\bibinfo {author} {\bibfnamefont {J.~T.}\ \bibnamefont {Barreiro}}, \bibinfo {author} {\bibfnamefont {T.-C.}\ \bibnamefont {Wei}},\ and\ \bibinfo {author} {\bibfnamefont {P.~G.}\ \bibnamefont {Kwiat}},\ }\bibfield  {title} {\bibinfo {title} {Beating the channel capacity limit for linear photonic superdense coding},\ }\href {https://doi.org/10.1038/nphys919} {\bibfield  {journal} {\bibinfo  {journal} {Nature Phys}\ }\textbf {\bibinfo {volume} {4}},\ \bibinfo {pages} {282} (\bibinfo {year} {2008})}\BibitemShut {NoStop}%
\bibitem [{\citenamefont {Yamazaki}\ \emph {et~al.}(2023{\natexlab{b}})\citenamefont {Yamazaki}, \citenamefont {Ikuta},\ and\ \citenamefont {Yamamoto}}]{Yamazaki_2023}%
  \BibitemOpen
  \bibfield  {author} {\bibinfo {author} {\bibfnamefont {T.}~\bibnamefont {Yamazaki}}, \bibinfo {author} {\bibfnamefont {R.}~\bibnamefont {Ikuta}},\ and\ \bibinfo {author} {\bibfnamefont {T.}~\bibnamefont {Yamamoto}},\ }\href@noop {} {\bibinfo {title} {Stabilizer formalism in linear optics and application to {{Bell-state}} discrimination}} (\bibinfo {year} {2023}{\natexlab{b}}),\ \Eprint {https://arxiv.org/abs/2301.06551} {arXiv:2301.06551} \BibitemShut {NoStop}%
\bibitem [{Note2()}]{Note2}%
  \BibitemOpen
  \bibinfo {note} {Each condition is written as $\Delta _{y} \pm \Delta _{z}=0 \protect \pmod {d}$ where $\Delta _{y}=y_1-y_0$ and $\Delta _{z}=z_1-z_0$. The both conditions are satisfied iff $2\Delta _y = 2\Delta _z = 0 \protect \pmod {d}$, that is, $\Delta _y = \Delta _z = d/2$, which imposes $d$ to be even}\BibitemShut {NoStop}%
\bibitem [{\citenamefont {Zhou}\ \emph {et~al.}(2015)\citenamefont {Zhou}, \citenamefont {Hua}, \citenamefont {Liu}, \citenamefont {Chen}, \citenamefont {Xu}, \citenamefont {Han}, \citenamefont {Li},\ and\ \citenamefont {Guo}}]{Zhou_Phys.Rev.Lett.2015}%
  \BibitemOpen
  \bibfield  {author} {\bibinfo {author} {\bibfnamefont {Z.-Q.}\ \bibnamefont {Zhou}}, \bibinfo {author} {\bibfnamefont {Y.-L.}\ \bibnamefont {Hua}}, \bibinfo {author} {\bibfnamefont {X.}~\bibnamefont {Liu}}, \bibinfo {author} {\bibfnamefont {G.}~\bibnamefont {Chen}}, \bibinfo {author} {\bibfnamefont {J.-S.}\ \bibnamefont {Xu}}, \bibinfo {author} {\bibfnamefont {Y.-J.}\ \bibnamefont {Han}}, \bibinfo {author} {\bibfnamefont {C.-F.}\ \bibnamefont {Li}},\ and\ \bibinfo {author} {\bibfnamefont {G.-C.}\ \bibnamefont {Guo}},\ }\bibfield  {title} {\bibinfo {title} {Quantum {{Storage}} of {{Three-Dimensional Orbital-Angular-Momentum Entanglement}} in a {{Crystal}}},\ }\href {https://doi.org/10.1103/PhysRevLett.115.070502} {\bibfield  {journal} {\bibinfo  {journal} {Phys. Rev. Lett.}\ }\textbf {\bibinfo {volume} {115}},\ \bibinfo {pages} {070502} (\bibinfo {year} {2015})}\BibitemShut {NoStop}%
\bibitem [{\citenamefont {Ding}\ \emph {et~al.}(2016)\citenamefont {Ding}, \citenamefont {Zhang}, \citenamefont {Shi}, \citenamefont {Zhou}, \citenamefont {Li}, \citenamefont {Shi},\ and\ \citenamefont {Guo}}]{Ding_LightSci.Appl.2016}%
  \BibitemOpen
  \bibfield  {author} {\bibinfo {author} {\bibfnamefont {D.-S.}\ \bibnamefont {Ding}}, \bibinfo {author} {\bibfnamefont {W.}~\bibnamefont {Zhang}}, \bibinfo {author} {\bibfnamefont {S.}~\bibnamefont {Shi}}, \bibinfo {author} {\bibfnamefont {Z.-Y.}\ \bibnamefont {Zhou}}, \bibinfo {author} {\bibfnamefont {Y.}~\bibnamefont {Li}}, \bibinfo {author} {\bibfnamefont {B.-S.}\ \bibnamefont {Shi}},\ and\ \bibinfo {author} {\bibfnamefont {G.-C.}\ \bibnamefont {Guo}},\ }\bibfield  {title} {\bibinfo {title} {High-dimensional entanglement between distant atomic-ensemble memories},\ }\href {https://doi.org/10.1038/lsa.2016.157} {\bibfield  {journal} {\bibinfo  {journal} {Light Sci Appl}\ }\textbf {\bibinfo {volume} {5}},\ \bibinfo {pages} {e16157} (\bibinfo {year} {2016})}\BibitemShut {NoStop}%
\bibitem [{\citenamefont {Zhang}\ \emph {et~al.}(2016)\citenamefont {Zhang}, \citenamefont {Ding}, \citenamefont {Dong}, \citenamefont {Shi}, \citenamefont {Wang}, \citenamefont {Liu}, \citenamefont {Li}, \citenamefont {Zhou}, \citenamefont {Shi},\ and\ \citenamefont {Guo}}]{Zhang_Nat.Commun.2016}%
  \BibitemOpen
  \bibfield  {author} {\bibinfo {author} {\bibfnamefont {W.}~\bibnamefont {Zhang}}, \bibinfo {author} {\bibfnamefont {D.-S.}\ \bibnamefont {Ding}}, \bibinfo {author} {\bibfnamefont {M.-X.}\ \bibnamefont {Dong}}, \bibinfo {author} {\bibfnamefont {S.}~\bibnamefont {Shi}}, \bibinfo {author} {\bibfnamefont {K.}~\bibnamefont {Wang}}, \bibinfo {author} {\bibfnamefont {S.-L.}\ \bibnamefont {Liu}}, \bibinfo {author} {\bibfnamefont {Y.}~\bibnamefont {Li}}, \bibinfo {author} {\bibfnamefont {Z.-Y.}\ \bibnamefont {Zhou}}, \bibinfo {author} {\bibfnamefont {B.-S.}\ \bibnamefont {Shi}},\ and\ \bibinfo {author} {\bibfnamefont {G.-C.}\ \bibnamefont {Guo}},\ }\bibfield  {title} {\bibinfo {title} {Experimental realization of entanglement in multiple degrees of freedom between two quantum memories},\ }\href {https://doi.org/10.1038/ncomms13514} {\bibfield  {journal} {\bibinfo  {journal} {Nat Commun}\ }\textbf {\bibinfo {volume} {7}},\ \bibinfo {pages} {13514} (\bibinfo {year} {2016})}\BibitemShut {NoStop}%
\bibitem [{\citenamefont {Tiranov}\ \emph {et~al.}(2017)\citenamefont {Tiranov}, \citenamefont {Designolle}, \citenamefont {Cruzeiro}, \citenamefont {Lavoie}, \citenamefont {Brunner}, \citenamefont {Afzelius}, \citenamefont {Huber},\ and\ \citenamefont {Gisin}}]{Tiranov_Phys.Rev.A2017}%
  \BibitemOpen
  \bibfield  {author} {\bibinfo {author} {\bibfnamefont {A.}~\bibnamefont {Tiranov}}, \bibinfo {author} {\bibfnamefont {S.}~\bibnamefont {Designolle}}, \bibinfo {author} {\bibfnamefont {E.~Z.}\ \bibnamefont {Cruzeiro}}, \bibinfo {author} {\bibfnamefont {J.}~\bibnamefont {Lavoie}}, \bibinfo {author} {\bibfnamefont {N.}~\bibnamefont {Brunner}}, \bibinfo {author} {\bibfnamefont {M.}~\bibnamefont {Afzelius}}, \bibinfo {author} {\bibfnamefont {M.}~\bibnamefont {Huber}},\ and\ \bibinfo {author} {\bibfnamefont {N.}~\bibnamefont {Gisin}},\ }\bibfield  {title} {\bibinfo {title} {Quantification of multidimensional entanglement stored in a crystal},\ }\href {https://doi.org/10.1103/PhysRevA.96.040303} {\bibfield  {journal} {\bibinfo  {journal} {Phys. Rev. A}\ }\textbf {\bibinfo {volume} {96}},\ \bibinfo {pages} {040303} (\bibinfo {year} {2017})}\BibitemShut {NoStop}%
\bibitem [{\citenamefont {Dong}\ \emph {et~al.}(2023)\citenamefont {Dong}, \citenamefont {Zhang}, \citenamefont {Zeng}, \citenamefont {Ye}, \citenamefont {Li}, \citenamefont {Guo}, \citenamefont {Ding},\ and\ \citenamefont {Shi}}]{Dong_Phys.Rev.Lett.2023}%
  \BibitemOpen
  \bibfield  {author} {\bibinfo {author} {\bibfnamefont {M.-X.}\ \bibnamefont {Dong}}, \bibinfo {author} {\bibfnamefont {W.-H.}\ \bibnamefont {Zhang}}, \bibinfo {author} {\bibfnamefont {L.}~\bibnamefont {Zeng}}, \bibinfo {author} {\bibfnamefont {Y.-H.}\ \bibnamefont {Ye}}, \bibinfo {author} {\bibfnamefont {D.-C.}\ \bibnamefont {Li}}, \bibinfo {author} {\bibfnamefont {G.-C.}\ \bibnamefont {Guo}}, \bibinfo {author} {\bibfnamefont {D.-S.}\ \bibnamefont {Ding}},\ and\ \bibinfo {author} {\bibfnamefont {B.-S.}\ \bibnamefont {Shi}},\ }\bibfield  {title} {\bibinfo {title} {Highly {{Efficient Storage}} of 25-{{Dimensional Photonic Qudit}} in a {{Cold-Atom-Based Quantum Memory}}},\ }\href {https://doi.org/10.1103/PhysRevLett.131.240801} {\bibfield  {journal} {\bibinfo  {journal} {Phys. Rev. Lett.}\ }\textbf {\bibinfo {volume} {131}},\ \bibinfo {pages} {240801} (\bibinfo {year} {2023})}\BibitemShut {NoStop}%
\bibitem [{\citenamefont {Xing}\ \emph {et~al.}(2023)\citenamefont {Xing}, \citenamefont {Hu}, \citenamefont {Guo}, \citenamefont {Liu}, \citenamefont {Li},\ and\ \citenamefont {Guo}}]{Xing_Opt.Express2023}%
  \BibitemOpen
  \bibfield  {author} {\bibinfo {author} {\bibfnamefont {W.-B.}\ \bibnamefont {Xing}}, \bibinfo {author} {\bibfnamefont {X.-M.}\ \bibnamefont {Hu}}, \bibinfo {author} {\bibfnamefont {Y.}~\bibnamefont {Guo}}, \bibinfo {author} {\bibfnamefont {B.-H.}\ \bibnamefont {Liu}}, \bibinfo {author} {\bibfnamefont {C.-F.}\ \bibnamefont {Li}},\ and\ \bibinfo {author} {\bibfnamefont {G.-C.}\ \bibnamefont {Guo}},\ }\bibfield  {title} {\bibinfo {title} {Preparation of multiphoton high-dimensional {{GHZ}} states},\ }\href {https://doi.org/10.1364/OE.494850} {\bibfield  {journal} {\bibinfo  {journal} {Opt. Express}\ }\textbf {\bibinfo {volume} {31}},\ \bibinfo {pages} {24887} (\bibinfo {year} {2023})}\BibitemShut {NoStop}%
\bibitem [{\citenamefont {Chin}\ \emph {et~al.}(2024)\citenamefont {Chin}, \citenamefont {Ryu},\ and\ \citenamefont {Kim}}]{Chin_2024}%
  \BibitemOpen
  \bibfield  {author} {\bibinfo {author} {\bibfnamefont {S.}~\bibnamefont {Chin}}, \bibinfo {author} {\bibfnamefont {J.}~\bibnamefont {Ryu}},\ and\ \bibinfo {author} {\bibfnamefont {Y.-S.}\ \bibnamefont {Kim}},\ }\href@noop {} {\bibinfo {title} {Exponentially {{Enhanced Scheme}} for the {{Heralded Qudit GHZ State}} in {{Linear Optics}}}} (\bibinfo {year} {2024}),\ \Eprint {https://arxiv.org/abs/2406.10830} {arXiv:2406.10830} \BibitemShut {NoStop}%
\bibitem [{\citenamefont {Munro}\ \emph {et~al.}(2015)\citenamefont {Munro}, \citenamefont {Azuma}, \citenamefont {Tamaki},\ and\ \citenamefont {Nemoto}}]{Munro_IEEEJ.Sel.Top.QuantumElectron.2015}%
  \BibitemOpen
  \bibfield  {author} {\bibinfo {author} {\bibfnamefont {W.~J.}\ \bibnamefont {Munro}}, \bibinfo {author} {\bibfnamefont {K.}~\bibnamefont {Azuma}}, \bibinfo {author} {\bibfnamefont {K.}~\bibnamefont {Tamaki}},\ and\ \bibinfo {author} {\bibfnamefont {K.}~\bibnamefont {Nemoto}},\ }\bibfield  {title} {\bibinfo {title} {Inside {{Quantum Repeaters}}},\ }\href {https://doi.org/10.1109/JSTQE.2015.2392076} {\bibfield  {journal} {\bibinfo  {journal} {IEEE Journal of Selected Topics in Quantum Electronics}\ }\textbf {\bibinfo {volume} {21}},\ \bibinfo {pages} {78} (\bibinfo {year} {2015})}\BibitemShut {NoStop}%
\bibitem [{\citenamefont {Muralidharan}\ \emph {et~al.}(2016)\citenamefont {Muralidharan}, \citenamefont {Li}, \citenamefont {Kim}, \citenamefont {L{\"u}tkenhaus}, \citenamefont {Lukin},\ and\ \citenamefont {Jiang}}]{Muralidharan_Sci.Rep.2016}%
  \BibitemOpen
  \bibfield  {author} {\bibinfo {author} {\bibfnamefont {S.}~\bibnamefont {Muralidharan}}, \bibinfo {author} {\bibfnamefont {L.}~\bibnamefont {Li}}, \bibinfo {author} {\bibfnamefont {J.}~\bibnamefont {Kim}}, \bibinfo {author} {\bibfnamefont {N.}~\bibnamefont {L{\"u}tkenhaus}}, \bibinfo {author} {\bibfnamefont {M.~D.}\ \bibnamefont {Lukin}},\ and\ \bibinfo {author} {\bibfnamefont {L.}~\bibnamefont {Jiang}},\ }\bibfield  {title} {\bibinfo {title} {Optimal architectures for long distance quantum communication},\ }\href {https://doi.org/10.1038/srep20463} {\bibfield  {journal} {\bibinfo  {journal} {Sci Rep}\ }\textbf {\bibinfo {volume} {6}},\ \bibinfo {pages} {20463} (\bibinfo {year} {2016})}\BibitemShut {NoStop}%
\bibitem [{\citenamefont {Azuma}\ \emph {et~al.}(2023)\citenamefont {Azuma}, \citenamefont {Economou}, \citenamefont {Elkouss}, \citenamefont {Hilaire}, \citenamefont {Jiang}, \citenamefont {Lo},\ and\ \citenamefont {Tzitrin}}]{Azuma_Rev.Mod.Phys.2023}%
  \BibitemOpen
  \bibfield  {author} {\bibinfo {author} {\bibfnamefont {K.}~\bibnamefont {Azuma}}, \bibinfo {author} {\bibfnamefont {S.~E.}\ \bibnamefont {Economou}}, \bibinfo {author} {\bibfnamefont {D.}~\bibnamefont {Elkouss}}, \bibinfo {author} {\bibfnamefont {P.}~\bibnamefont {Hilaire}}, \bibinfo {author} {\bibfnamefont {L.}~\bibnamefont {Jiang}}, \bibinfo {author} {\bibfnamefont {H.-K.}\ \bibnamefont {Lo}},\ and\ \bibinfo {author} {\bibfnamefont {I.}~\bibnamefont {Tzitrin}},\ }\bibfield  {title} {\bibinfo {title} {Quantum repeaters: {{From}} quantum networks to the quantum internet},\ }\href {https://doi.org/10.1103/RevModPhys.95.045006} {\bibfield  {journal} {\bibinfo  {journal} {Rev. Mod. Phys.}\ }\textbf {\bibinfo {volume} {95}},\ \bibinfo {pages} {045006} (\bibinfo {year} {2023})}\BibitemShut {NoStop}%
\bibitem [{\citenamefont {Briegel}\ \emph {et~al.}(1998)\citenamefont {Briegel}, \citenamefont {D{\"u}r}, \citenamefont {Cirac},\ and\ \citenamefont {Zoller}}]{Briegel_Phys.Rev.Lett.1998a}%
  \BibitemOpen
  \bibfield  {author} {\bibinfo {author} {\bibfnamefont {H.-J.}\ \bibnamefont {Briegel}}, \bibinfo {author} {\bibfnamefont {W.}~\bibnamefont {D{\"u}r}}, \bibinfo {author} {\bibfnamefont {J.~I.}\ \bibnamefont {Cirac}},\ and\ \bibinfo {author} {\bibfnamefont {P.}~\bibnamefont {Zoller}},\ }\bibfield  {title} {\bibinfo {title} {Quantum {{Repeaters}}: {{The Role}} of {{Imperfect Local Operations}} in {{Quantum Communication}}},\ }\href {https://doi.org/10.1103/PhysRevLett.81.5932} {\bibfield  {journal} {\bibinfo  {journal} {Phys. Rev. Lett.}\ }\textbf {\bibinfo {volume} {81}},\ \bibinfo {pages} {5932} (\bibinfo {year} {1998})}\BibitemShut {NoStop}%
\bibitem [{\citenamefont {Duan}\ \emph {et~al.}(2001)\citenamefont {Duan}, \citenamefont {Lukin}, \citenamefont {Cirac},\ and\ \citenamefont {Zoller}}]{Duan_Nature2001}%
  \BibitemOpen
  \bibfield  {author} {\bibinfo {author} {\bibfnamefont {L.~M.}\ \bibnamefont {Duan}}, \bibinfo {author} {\bibfnamefont {M.~D.}\ \bibnamefont {Lukin}}, \bibinfo {author} {\bibfnamefont {J.~I.}\ \bibnamefont {Cirac}},\ and\ \bibinfo {author} {\bibfnamefont {P.}~\bibnamefont {Zoller}},\ }\bibfield  {title} {\bibinfo {title} {Long-distance quantum communication with atomic ensembles and linear optics},\ }\href {https://doi.org/10.1038/35106500} {\bibfield  {journal} {\bibinfo  {journal} {Nature}\ }\textbf {\bibinfo {volume} {414}},\ \bibinfo {pages} {413} (\bibinfo {year} {2001})}\BibitemShut {NoStop}%
\bibitem [{\citenamefont {Sangouard}\ \emph {et~al.}(2011)\citenamefont {Sangouard}, \citenamefont {Simon}, \citenamefont {{de Riedmatten}},\ and\ \citenamefont {Gisin}}]{Sangouard_Rev.Mod.Phys.2011}%
  \BibitemOpen
  \bibfield  {author} {\bibinfo {author} {\bibfnamefont {N.}~\bibnamefont {Sangouard}}, \bibinfo {author} {\bibfnamefont {C.}~\bibnamefont {Simon}}, \bibinfo {author} {\bibfnamefont {H.}~\bibnamefont {{de Riedmatten}}},\ and\ \bibinfo {author} {\bibfnamefont {N.}~\bibnamefont {Gisin}},\ }\bibfield  {title} {\bibinfo {title} {Quantum repeaters based on atomic ensembles and linear optics},\ }\href {https://doi.org/10.1103/RevModPhys.83.33} {\bibfield  {journal} {\bibinfo  {journal} {Rev. Mod. Phys.}\ }\textbf {\bibinfo {volume} {83}},\ \bibinfo {pages} {33} (\bibinfo {year} {2011})}\BibitemShut {NoStop}%
\bibitem [{\citenamefont {Jiang}\ \emph {et~al.}(2009)\citenamefont {Jiang}, \citenamefont {Taylor}, \citenamefont {Nemoto}, \citenamefont {Munro}, \citenamefont {Van~Meter},\ and\ \citenamefont {Lukin}}]{Jiang_Phys.Rev.A2009}%
  \BibitemOpen
  \bibfield  {author} {\bibinfo {author} {\bibfnamefont {L.}~\bibnamefont {Jiang}}, \bibinfo {author} {\bibfnamefont {J.~M.}\ \bibnamefont {Taylor}}, \bibinfo {author} {\bibfnamefont {K.}~\bibnamefont {Nemoto}}, \bibinfo {author} {\bibfnamefont {W.~J.}\ \bibnamefont {Munro}}, \bibinfo {author} {\bibfnamefont {R.}~\bibnamefont {Van~Meter}},\ and\ \bibinfo {author} {\bibfnamefont {M.~D.}\ \bibnamefont {Lukin}},\ }\bibfield  {title} {\bibinfo {title} {Quantum repeater with encoding},\ }\href {https://doi.org/10.1103/PhysRevA.79.032325} {\bibfield  {journal} {\bibinfo  {journal} {Phys. Rev. A}\ }\textbf {\bibinfo {volume} {79}},\ \bibinfo {pages} {032325} (\bibinfo {year} {2009})}\BibitemShut {NoStop}%
\bibitem [{\citenamefont {Munro}\ \emph {et~al.}(2010)\citenamefont {Munro}, \citenamefont {Harrison}, \citenamefont {Stephens}, \citenamefont {Devitt},\ and\ \citenamefont {Nemoto}}]{Munro_Nat.Photonics2010}%
  \BibitemOpen
  \bibfield  {author} {\bibinfo {author} {\bibfnamefont {W.~J.}\ \bibnamefont {Munro}}, \bibinfo {author} {\bibfnamefont {K.~A.}\ \bibnamefont {Harrison}}, \bibinfo {author} {\bibfnamefont {A.~M.}\ \bibnamefont {Stephens}}, \bibinfo {author} {\bibfnamefont {S.~J.}\ \bibnamefont {Devitt}},\ and\ \bibinfo {author} {\bibfnamefont {K.}~\bibnamefont {Nemoto}},\ }\bibfield  {title} {\bibinfo {title} {From quantum multiplexing to high-performance quantum networking},\ }\href {https://doi.org/10.1038/nphoton.2010.213} {\bibfield  {journal} {\bibinfo  {journal} {Nature Photon}\ }\textbf {\bibinfo {volume} {4}},\ \bibinfo {pages} {792} (\bibinfo {year} {2010})}\BibitemShut {NoStop}%
\bibitem [{\citenamefont {Li}\ \emph {et~al.}(2013)\citenamefont {Li}, \citenamefont {Barrett}, \citenamefont {Stace},\ and\ \citenamefont {Benjamin}}]{Li_NewJ.Phys.2013}%
  \BibitemOpen
  \bibfield  {author} {\bibinfo {author} {\bibfnamefont {Y.}~\bibnamefont {Li}}, \bibinfo {author} {\bibfnamefont {S.~D.}\ \bibnamefont {Barrett}}, \bibinfo {author} {\bibfnamefont {T.~M.}\ \bibnamefont {Stace}},\ and\ \bibinfo {author} {\bibfnamefont {S.~C.}\ \bibnamefont {Benjamin}},\ }\bibfield  {title} {\bibinfo {title} {Long range failure-tolerant entanglement distribution},\ }\href {https://doi.org/10.1088/1367-2630/15/2/023012} {\bibfield  {journal} {\bibinfo  {journal} {New J. Phys.}\ }\textbf {\bibinfo {volume} {15}},\ \bibinfo {pages} {023012} (\bibinfo {year} {2013})}\BibitemShut {NoStop}%
\bibitem [{\citenamefont {Sangouard}\ \emph {et~al.}(2009)\citenamefont {Sangouard}, \citenamefont {Dubessy},\ and\ \citenamefont {Simon}}]{Sangouard_Phys.Rev.A2009}%
  \BibitemOpen
  \bibfield  {author} {\bibinfo {author} {\bibfnamefont {N.}~\bibnamefont {Sangouard}}, \bibinfo {author} {\bibfnamefont {R.}~\bibnamefont {Dubessy}},\ and\ \bibinfo {author} {\bibfnamefont {C.}~\bibnamefont {Simon}},\ }\bibfield  {title} {\bibinfo {title} {Quantum repeaters based on single trapped ions},\ }\href {https://doi.org/10.1103/PhysRevA.79.042340} {\bibfield  {journal} {\bibinfo  {journal} {Phys. Rev. A}\ }\textbf {\bibinfo {volume} {79}},\ \bibinfo {pages} {042340} (\bibinfo {year} {2009})}\BibitemShut {NoStop}%
\bibitem [{\citenamefont {Asadi}\ \emph {et~al.}(2018)\citenamefont {Asadi}, \citenamefont {Lauk}, \citenamefont {Wein}, \citenamefont {Sinclair}, \citenamefont {O'Brien},\ and\ \citenamefont {Simon}}]{Asadi_Quantum2018}%
  \BibitemOpen
  \bibfield  {author} {\bibinfo {author} {\bibfnamefont {F.~K.}\ \bibnamefont {Asadi}}, \bibinfo {author} {\bibfnamefont {N.}~\bibnamefont {Lauk}}, \bibinfo {author} {\bibfnamefont {S.}~\bibnamefont {Wein}}, \bibinfo {author} {\bibfnamefont {N.}~\bibnamefont {Sinclair}}, \bibinfo {author} {\bibfnamefont {C.}~\bibnamefont {O'Brien}},\ and\ \bibinfo {author} {\bibfnamefont {C.}~\bibnamefont {Simon}},\ }\bibfield  {title} {\bibinfo {title} {Quantum repeaters with individual rare-earth ions at telecommunication wavelengths},\ }\href {https://doi.org/10.22331/q-2018-09-13-93} {\bibfield  {journal} {\bibinfo  {journal} {Quantum}\ }\textbf {\bibinfo {volume} {2}},\ \bibinfo {pages} {93} (\bibinfo {year} {2018})}\BibitemShut {NoStop}%
\bibitem [{\citenamefont {Asadi}\ \emph {et~al.}(2020)\citenamefont {Asadi}, \citenamefont {Wein},\ and\ \citenamefont {Simon}}]{Asadi_QuantumSci.Technol.2020}%
  \BibitemOpen
  \bibfield  {author} {\bibinfo {author} {\bibfnamefont {F.~K.}\ \bibnamefont {Asadi}}, \bibinfo {author} {\bibfnamefont {S.~C.}\ \bibnamefont {Wein}},\ and\ \bibinfo {author} {\bibfnamefont {C.}~\bibnamefont {Simon}},\ }\bibfield  {title} {\bibinfo {title} {Protocols for long-distance quantum communication with single {{167Er}} ions},\ }\href {https://doi.org/10.1088/2058-9565/abae7c} {\bibfield  {journal} {\bibinfo  {journal} {Quantum Sci. Technol.}\ }\textbf {\bibinfo {volume} {5}},\ \bibinfo {pages} {045015} (\bibinfo {year} {2020})}\BibitemShut {NoStop}%
\bibitem [{\citenamefont {Lindner}\ and\ \citenamefont {Rudolph}(2009)}]{Lindner_Phys.Rev.Lett.2009}%
  \BibitemOpen
  \bibfield  {author} {\bibinfo {author} {\bibfnamefont {N.~H.}\ \bibnamefont {Lindner}}\ and\ \bibinfo {author} {\bibfnamefont {T.}~\bibnamefont {Rudolph}},\ }\bibfield  {title} {\bibinfo {title} {Proposal for pulsed on-demand sources of photonic cluster state strings},\ }\href {https://doi.org/10.1103/PhysRevLett.103.113602} {\bibfield  {journal} {\bibinfo  {journal} {Phys. Rev. Lett.}\ }\textbf {\bibinfo {volume} {103}},\ \bibinfo {pages} {113602} (\bibinfo {year} {2009})}\BibitemShut {NoStop}%
\bibitem [{\citenamefont {Schwartz}\ \emph {et~al.}(2016)\citenamefont {Schwartz}, \citenamefont {Cogan}, \citenamefont {Schmidgall}, \citenamefont {Don}, \citenamefont {Gantz}, \citenamefont {Kenneth}, \citenamefont {Lindner},\ and\ \citenamefont {Gershoni}}]{Schwartz_Science2016}%
  \BibitemOpen
  \bibfield  {author} {\bibinfo {author} {\bibfnamefont {I.}~\bibnamefont {Schwartz}}, \bibinfo {author} {\bibfnamefont {D.}~\bibnamefont {Cogan}}, \bibinfo {author} {\bibfnamefont {E.~R.}\ \bibnamefont {Schmidgall}}, \bibinfo {author} {\bibfnamefont {Y.}~\bibnamefont {Don}}, \bibinfo {author} {\bibfnamefont {L.}~\bibnamefont {Gantz}}, \bibinfo {author} {\bibfnamefont {O.}~\bibnamefont {Kenneth}}, \bibinfo {author} {\bibfnamefont {N.~H.}\ \bibnamefont {Lindner}},\ and\ \bibinfo {author} {\bibfnamefont {D.}~\bibnamefont {Gershoni}},\ }\bibfield  {title} {\bibinfo {title} {Deterministic generation of a cluster state of entangled photons},\ }\href {https://doi.org/10.1126/science.aah4758} {\bibfield  {journal} {\bibinfo  {journal} {Science}\ }\textbf {\bibinfo {volume} {354}},\ \bibinfo {pages} {434} (\bibinfo {year} {2016})}\BibitemShut {NoStop}%
\bibitem [{\citenamefont {Bell}\ \emph {et~al.}(2022)\citenamefont {Bell}, \citenamefont {Bulmer}, \citenamefont {Jones}, \citenamefont {Paesani}, \citenamefont {McCutcheon},\ and\ \citenamefont {Laing}}]{Bell_NewJ.Phys.2022}%
  \BibitemOpen
  \bibfield  {author} {\bibinfo {author} {\bibfnamefont {T.~J.}\ \bibnamefont {Bell}}, \bibinfo {author} {\bibfnamefont {J.~F.~F.}\ \bibnamefont {Bulmer}}, \bibinfo {author} {\bibfnamefont {A.~E.}\ \bibnamefont {Jones}}, \bibinfo {author} {\bibfnamefont {S.}~\bibnamefont {Paesani}}, \bibinfo {author} {\bibfnamefont {D.~P.~S.}\ \bibnamefont {McCutcheon}},\ and\ \bibinfo {author} {\bibfnamefont {A.}~\bibnamefont {Laing}},\ }\bibfield  {title} {\bibinfo {title} {Protocol for generation of high-dimensional entanglement from an array of non-interacting photon emitters},\ }\href {https://doi.org/10.1088/1367-2630/ac475d} {\bibfield  {journal} {\bibinfo  {journal} {New J. Phys.}\ }\textbf {\bibinfo {volume} {24}},\ \bibinfo {pages} {013032} (\bibinfo {year} {2022})}\BibitemShut {NoStop}%
\bibitem [{\citenamefont {Raissi}\ \emph {et~al.}(2024)\citenamefont {Raissi}, \citenamefont {Barnes},\ and\ \citenamefont {Economou}}]{Raissi_PRXQuantum2024}%
  \BibitemOpen
  \bibfield  {author} {\bibinfo {author} {\bibfnamefont {Z.}~\bibnamefont {Raissi}}, \bibinfo {author} {\bibfnamefont {E.}~\bibnamefont {Barnes}},\ and\ \bibinfo {author} {\bibfnamefont {S.~E.}\ \bibnamefont {Economou}},\ }\bibfield  {title} {\bibinfo {title} {Deterministic {{Generation}} of {{Qudit Photonic Graph States}} from {{Quantum Emitters}}},\ }\href {https://doi.org/10.1103/PRXQuantum.5.020346} {\bibfield  {journal} {\bibinfo  {journal} {PRX Quantum}\ }\textbf {\bibinfo {volume} {5}},\ \bibinfo {pages} {020346} (\bibinfo {year} {2024})}\BibitemShut {NoStop}%
\end{thebibliography}%


\begin{thebibliography}{10}%
\makeatletter
\providecommand \@ifxundefined [1]{%
 \@ifx{#1\undefined}
}%
\providecommand \@ifnum [1]{%
 \ifnum #1\expandafter \@firstoftwo
 \else \expandafter \@secondoftwo
 \fi
}%
\providecommand \@ifx [1]{%
 \ifx #1\expandafter \@firstoftwo
 \else \expandafter \@secondoftwo
 \fi
}%
\providecommand \natexlab [1]{#1}%
\providecommand \enquote  [1]{``#1''}%
\providecommand \bibnamefont  [1]{#1}%
\providecommand \bibfnamefont [1]{#1}%
\providecommand \citenamefont [1]{#1}%
\providecommand \href@noop [0]{\@secondoftwo}%
\providecommand \href [0]{\begingroup \@sanitize@url \@href}%
\providecommand \@href[1]{\@@startlink{#1}\@@href}%
\providecommand \@@href[1]{\endgroup#1\@@endlink}%
\providecommand \@sanitize@url [0]{\catcode `\\12\catcode `\$12\catcode `\&12\catcode `\#12\catcode `\^12\catcode `\_12\catcode `\%12\relax}%
\providecommand \@@startlink[1]{}%
\providecommand \@@endlink[0]{}%
\providecommand \url  [0]{\begingroup\@sanitize@url \@url }%
\providecommand \@url [1]{\endgroup\@href {#1}{\urlprefix }}%
\providecommand \urlprefix  [0]{URL }%
\providecommand \Eprint [0]{\href }%
\providecommand \doibase [0]{https://doi.org/}%
\providecommand \selectlanguage [0]{\@gobble}%
\providecommand \bibinfo  [0]{\@secondoftwo}%
\providecommand \bibfield  [0]{\@secondoftwo}%
\providecommand \translation [1]{[#1]}%
\providecommand \BibitemOpen [0]{}%
\providecommand \bibitemStop [0]{}%
\providecommand \bibitemNoStop [0]{.\EOS\space}%
\providecommand \EOS [0]{\spacefactor3000\relax}%
\providecommand \BibitemShut  [1]{\csname bibitem#1\endcsname}%
\let\auto@bib@innerbib\@empty
\bibitem [{\citenamefont {Yamazaki}\ \emph {et~al.}(2023)\citenamefont {Yamazaki}, \citenamefont {Ikuta},\ and\ \citenamefont {Yamamoto}}]{Yamazaki_2023}%
  \BibitemOpen
  \bibfield  {author} {\bibinfo {author} {\bibfnamefont {T.}~\bibnamefont {Yamazaki}}, \bibinfo {author} {\bibfnamefont {R.}~\bibnamefont {Ikuta}},\ and\ \bibinfo {author} {\bibfnamefont {T.}~\bibnamefont {Yamamoto}},\ }\href@noop {} {\bibinfo {title} {Stabilizer formalism in linear optics and application to {{Bell-state}} discrimination}} (\bibinfo {year} {2023}),\ \Eprint {https://arxiv.org/abs/2301.06551} {arXiv:2301.06551} \BibitemShut {NoStop}%
\bibitem [{Note1()}]{Note1}%
  \BibitemOpen
  \bibinfo {note} {In the bosonic stabilizer formalism~\cite {Yamazaki_2023}, we say the LOC used in the Lemma as \protect \textit {the measurement of stabilizer $I_2^{\otimes k-p} \otimes X \otimes I_2^{\otimes p}\otimes I_d$} and represent obtaining one of measurement patterns included in $S_\pm $ with \protect \textit {the measurement result of $I_2^{\otimes k-p} \otimes X \otimes I_2^{\otimes p}\otimes I_d=\pm 1$}.}\BibitemShut {Stop}%
\bibitem [{\citenamefont {Sangouard}\ \emph {et~al.}(2011)\citenamefont {Sangouard}, \citenamefont {Simon}, \citenamefont {{de Riedmatten}},\ and\ \citenamefont {Gisin}}]{Sangouard_Rev.Mod.Phys.2011}%
  \BibitemOpen
  \bibfield  {author} {\bibinfo {author} {\bibfnamefont {N.}~\bibnamefont {Sangouard}}, \bibinfo {author} {\bibfnamefont {C.}~\bibnamefont {Simon}}, \bibinfo {author} {\bibfnamefont {H.}~\bibnamefont {{de Riedmatten}}},\ and\ \bibinfo {author} {\bibfnamefont {N.}~\bibnamefont {Gisin}},\ }\bibfield  {title} {\bibinfo {title} {Quantum repeaters based on atomic ensembles and linear optics},\ }\href {https://doi.org/10.1103/RevModPhys.83.33} {\bibfield  {journal} {\bibinfo  {journal} {Rev. Mod. Phys.}\ }\textbf {\bibinfo {volume} {83}},\ \bibinfo {pages} {33} (\bibinfo {year} {2011})}\BibitemShut {NoStop}%
\bibitem [{Note2()}]{Note2}%
  \BibitemOpen
  \bibinfo {note} {This heralding signal is necessary even when the $k$th level of entanglement swapping failed, because the $(2^{k+1}i-2^k)$th repeater node cannot discard the photons preserved in the quantum memories and restart the entanglement generation step until it knows that the $k$th level of entanglement swapping failed}\BibitemShut {NoStop}%
\bibitem [{Note10()}]{Note10}%
  \BibitemOpen
  \bibinfo {note} {The heralding signal to announce the result of the last entanglement swapping is also necessary for the synchronization among all the repeater node. In principle, such synchronization is unnecessary because each repeater node can start the next entanglement generation step immediately after performing the entanglement swapping although the entanglement generation will never succeed until the adjacent node also finishes the entanglement swapping and starts the entanglement generation step. However, the absence of the synchronization causes a difference in the timing for the entanglement generation step to start among the repeater nodes, which in turn can make, for example, the required memory time longer. Here, for simplicity, we ignore such effects although the transmission time of the heralding signal for the synchronization is not included in the cycle time of the protocol}\BibitemShut {NoStop}%
\bibitem [{Note3()}]{Note3}%
  \BibitemOpen
  \bibinfo {note} {$c'T_\protect \text {first}(n)/L_\protect \text {tot}$ is lower bounded by $2\alpha ^2/3$ for $n>1$ and $\alpha /P_\protect \text {g}(L_\protect \text {tot}/2)$ for $n=1$. Thus, it is not only upper bounded but also lower bounded by a constant with respect to $L_\protect \text {tot}$.}\BibitemShut {Stop}%
\bibitem [{\citenamefont {Azuma}\ \emph {et~al.}(2021)\citenamefont {Azuma}, \citenamefont {B{\"a}uml}, \citenamefont {Coopmans}, \citenamefont {Elkouss},\ and\ \citenamefont {Li}}]{Azuma_AVSQuantumSci.2021a}%
  \BibitemOpen
  \bibfield  {author} {\bibinfo {author} {\bibfnamefont {K.}~\bibnamefont {Azuma}}, \bibinfo {author} {\bibfnamefont {S.}~\bibnamefont {B{\"a}uml}}, \bibinfo {author} {\bibfnamefont {T.}~\bibnamefont {Coopmans}}, \bibinfo {author} {\bibfnamefont {D.}~\bibnamefont {Elkouss}},\ and\ \bibinfo {author} {\bibfnamefont {B.}~\bibnamefont {Li}},\ }\bibfield  {title} {\bibinfo {title} {Tools for quantum network design},\ }\href {https://doi.org/10.1116/5.0024062} {\bibfield  {journal} {\bibinfo  {journal} {AVS Quantum Science}\ }\textbf {\bibinfo {volume} {3}},\ \bibinfo {pages} {014101} (\bibinfo {year} {2021})}\BibitemShut {NoStop}%
\bibitem [{\citenamefont {Jiang}\ \emph {et~al.}(2009)\citenamefont {Jiang}, \citenamefont {Taylor}, \citenamefont {Nemoto}, \citenamefont {Munro}, \citenamefont {Van~Meter},\ and\ \citenamefont {Lukin}}]{Jiang_Phys.Rev.A2009}%
  \BibitemOpen
  \bibfield  {author} {\bibinfo {author} {\bibfnamefont {L.}~\bibnamefont {Jiang}}, \bibinfo {author} {\bibfnamefont {J.~M.}\ \bibnamefont {Taylor}}, \bibinfo {author} {\bibfnamefont {K.}~\bibnamefont {Nemoto}}, \bibinfo {author} {\bibfnamefont {W.~J.}\ \bibnamefont {Munro}}, \bibinfo {author} {\bibfnamefont {R.}~\bibnamefont {Van~Meter}},\ and\ \bibinfo {author} {\bibfnamefont {M.~D.}\ \bibnamefont {Lukin}},\ }\bibfield  {title} {\bibinfo {title} {Quantum repeater with encoding},\ }\href {https://doi.org/10.1103/PhysRevA.79.032325} {\bibfield  {journal} {\bibinfo  {journal} {Phys. Rev. A}\ }\textbf {\bibinfo {volume} {79}},\ \bibinfo {pages} {032325} (\bibinfo {year} {2009})}\BibitemShut {NoStop}%
\bibitem [{\citenamefont {Munro}\ \emph {et~al.}(2010)\citenamefont {Munro}, \citenamefont {Harrison}, \citenamefont {Stephens}, \citenamefont {Devitt},\ and\ \citenamefont {Nemoto}}]{Munro_Nat.Photonics2010}%
  \BibitemOpen
  \bibfield  {author} {\bibinfo {author} {\bibfnamefont {W.~J.}\ \bibnamefont {Munro}}, \bibinfo {author} {\bibfnamefont {K.~A.}\ \bibnamefont {Harrison}}, \bibinfo {author} {\bibfnamefont {A.~M.}\ \bibnamefont {Stephens}}, \bibinfo {author} {\bibfnamefont {S.~J.}\ \bibnamefont {Devitt}},\ and\ \bibinfo {author} {\bibfnamefont {K.}~\bibnamefont {Nemoto}},\ }\bibfield  {title} {\bibinfo {title} {From quantum multiplexing to high-performance quantum networking},\ }\href {https://doi.org/10.1038/nphoton.2010.213} {\bibfield  {journal} {\bibinfo  {journal} {Nature Photon}\ }\textbf {\bibinfo {volume} {4}},\ \bibinfo {pages} {792} (\bibinfo {year} {2010})}\BibitemShut {NoStop}%
\bibitem [{\citenamefont {Li}\ \emph {et~al.}(2013)\citenamefont {Li}, \citenamefont {Barrett}, \citenamefont {Stace},\ and\ \citenamefont {Benjamin}}]{Li_NewJ.Phys.2013}%
  \BibitemOpen
  \bibfield  {author} {\bibinfo {author} {\bibfnamefont {Y.}~\bibnamefont {Li}}, \bibinfo {author} {\bibfnamefont {S.~D.}\ \bibnamefont {Barrett}}, \bibinfo {author} {\bibfnamefont {T.~M.}\ \bibnamefont {Stace}},\ and\ \bibinfo {author} {\bibfnamefont {S.~C.}\ \bibnamefont {Benjamin}},\ }\bibfield  {title} {\bibinfo {title} {Long range failure-tolerant entanglement distribution},\ }\href {https://doi.org/10.1088/1367-2630/15/2/023012} {\bibfield  {journal} {\bibinfo  {journal} {New J. Phys.}\ }\textbf {\bibinfo {volume} {15}},\ \bibinfo {pages} {023012} (\bibinfo {year} {2013})}\BibitemShut {NoStop}%
\end{thebibliography}%

\end{document}



\section{Supplemental Material}
\subsection{A. Analysis of linear-optical circuits based on bosonic stabilizer formalism}
\begin{figure*}[bp]
    \begin{minipage}[b]{0.65\linewidth}
    \centering
    \subcaption{}\label{fig:graphical}
    \includegraphics[scale=0.8, trim=0 0 0 0, clip]{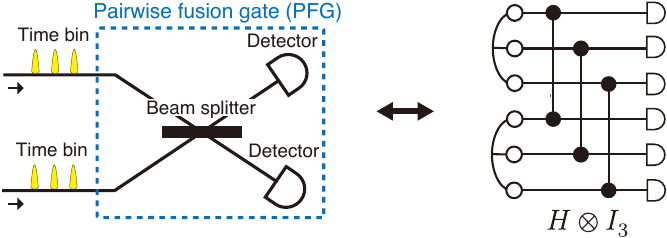}
    \subcaption{}\label{fig:graphical_ancilla}
    \hspace{-10pt}
    \vspace{-5pt}
    \includegraphics[scale=0.67, trim=0 0 0 0, clip]{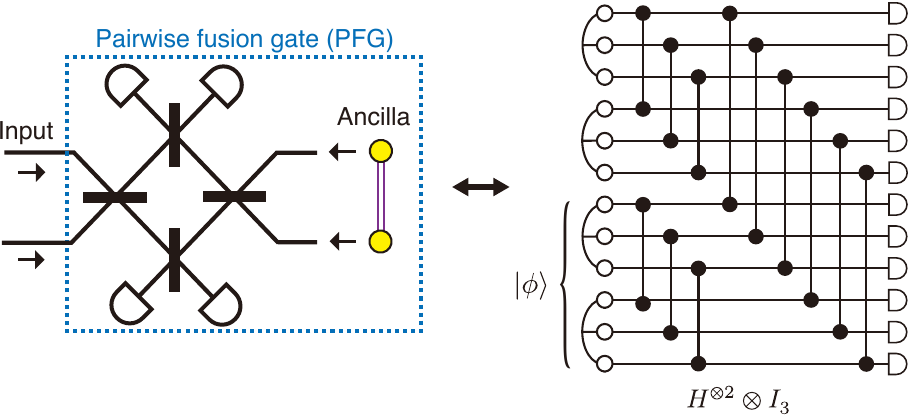}
    \end{minipage}
    \begin{minipage}[b]{0.34\linewidth}
    \centering
    \subcaption{}\label{fig:second_order}
    \hspace{-20pt}
    \includegraphics[scale=0.61, trim=0 0 0 0, clip]{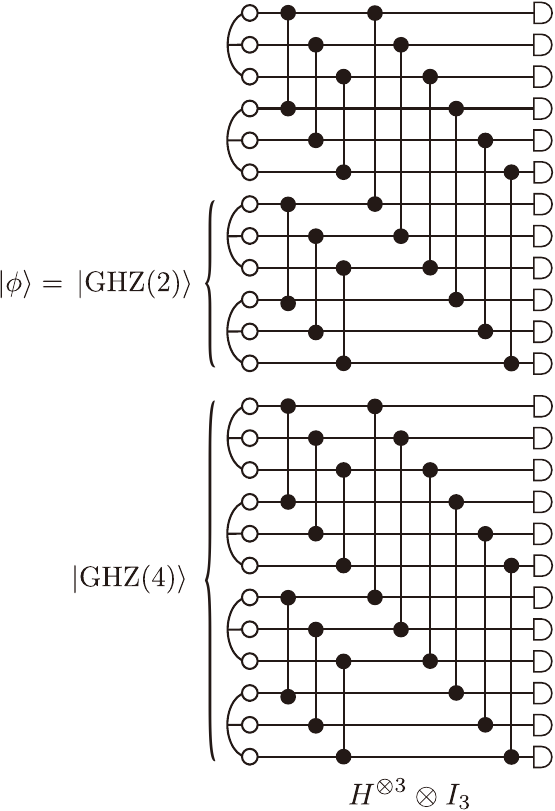}
    \end{minipage}
    \caption{Graphical representation of the proposed linear-optical circuits (LOCs) for $d=3$ and their correspondence with the experimental setups for time-bin qudits in the manuscript.\ (a) The LOC without ancilla photons ($k=0$).\ (b) The LOC with a Bell state as the ancilla ($k=1$).\ (c) The LOC with a Bell state and a four-qudit GHZ state as the ancilla ($k=2$).}
\end{figure*}

Here, we identify the quantum measurements on two qudits to which the linear optical circuits (LOCs) proposed in the manuscript corresponds, based on the \textit{bosonic stabilizer formalism}~\cite{Yamazaki_2023}.
First, we introduce a graphical representation of LOCs, instead of showing experimental setups especially for the case of the time-bin encoding.
The modes are represented by horizontal lines, which correspond to the modes indexed by $0, 1, \dots$ from the top to the bottom. 
Photon-number-resolving detectors are represented by deformed semicircles~$\includegraphics[scale=0.9, trim=0 0 0 0, clip,valign=c]{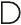}$.
The LOC corresponding to the transformation on two modes specified by transfer matrix $H$ is represented by two black points on the modes, connected with a vertical edge, such as 
\begin{equation}
    H \otimes I_2 = \includegraphics[scale=0.7, trim=0 0 0 0, clip,valign=c]{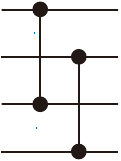},
\end{equation}
where $I_d$ is the $d\times d$ identity matrix.
We also explicitly represent the part of modes that constitutes a qudit in $d$-rail encoding by connecting white circles at the left ends of the horizontal lines and, if an input state of a part of modes is fixed, we write down the state of the corresponding qudits. For instance,
\begin{equation}
    \ket{\nu_0} = \includegraphics[scale=0.9, trim=0 0 0 0, clip, valign=c]{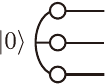}\notag
\end{equation}
for $d=3$.
Figures~\ref{fig:graphical} and \ref{fig:graphical_ancilla} represent the correspondence between the introduced graphical representations and the experimental setups for time-bin qudits shown in Fig.~\ref{fig:timebin_setup} and Fig.~\ref{fig:timebin_setup_ancilla} for $d=3$, respectively.

\subsubsection{Bosonic stabilizer formalism}
Here, we shortly introduce the bosonic stabilizer formalism~\cite{Yamazaki_2023}.
A time evolution $\mathcal{U}$ of Fock states induced by a combination of linear-optical elements is determined by a transformation $\mathcal{U}a_i^\dag \mathcal{U}^\dag = \sum_{j=0}^{m-1} a^\dag_j u_{ji}$ of creation operators $\{a_i^\dag\}_{i=0,\cdots,m-1}$, where transfer matrix $U$ is a unitary matrix.
Conversely, there is a map $B$ from a transfer matrix $U$ to the corresponding unitary operator $\mathcal{U}$ with $\mathcal{U} = B(U)$.
That is, for an input state $\ket*{\Psi_\text{in}}$ and the LOC consisting only of linear-optical elements, specified by a transfer matrix $U$, the output state is represented with this map $B$ as $\ket*{\Psi_\text{out}} = B(U)\ket*{\Psi_\text{in}}$.
From the definition, it is implied that
\begin{gather}
    B(UV)=B(U)B(V) \label{eq:homomorphism_product},\\
    B(U\oplus V)= B\qty(\mqty[U & 0 \\ 0 & V]) =  B(U)\otimes B(V) \label{eq:homomorphism_tensor}.
\end{gather}
For example, the LOCs represented by the transfer matrices of the Pauli-X and Pauli-Z matrices act on two-mode Fock states as
\begin{equation}\label{boson_Pauli}
    B(X)\ket*{n_0 n_1} = \ket*{n_1 n_0}, \quad
    B(Z)\ket*{n_0 n_1} = (-1)^{n_1} \ket*{n_0 n_1},
\end{equation}
respectively.
For a permutation matrix $P$ such as $X$, $B(P)$ corresponds to a permutation of modes.
For a diagonal matrix $D$ such as $Z$, $B(D)$ corresponds to a set of phase shifters.
Every Fock state $\ket*{\vb*{n}}=\ket*{n_0,n_1,\cdots}$ is an eigenstate of $B(D)$, and conversely, the eigenspace of $B(D)$ with eigenvalue $\lambda$ is spanned by the Fock states $\{\ket*{\vb*{n}}\}$ satisfying $B(D)\ket*{\vb*{n}}=\lambda \ket*{\vb*{n}}$.
From these consideration, we have the following lemma.
\begin{lemma}{}\label{lemma}
    Let states $\ket*{\Psi_+}$ and $\ket*{\Psi_-}$ in the $2^{k+1} d$-mode Fock space satisfy 
    \begin{equation}
        B(I_2^{\otimes k-p} \otimes X \otimes I_2^{\otimes p}\otimes I_d)\ket*{\Psi_\pm} = \pm \ket*{\Psi_\pm}\label{eq:symmetry_condition},
    \end{equation}
    respectively, for an integer $p \in \{0,\dots, k\}$.
    Then, these states can be distinguished by the LOC consisting of the interferometric part specified by transfer matrix $H^{\otimes k+1} \otimes I_d$ and $2^{k+1} d$ detectors~\footnote{In the bosonic stabilizer formalism~\cite{Yamazaki_2023}, we say the LOC used in the Lemma as \textit{the measurement of stabilizer $I_2^{\otimes k-p} \otimes X \otimes I_2^{\otimes p}\otimes I_d$} and represent obtaining one of measurement patterns included in $S_\pm$ with \textit{the measurement result of $I_2^{\otimes k-p} \otimes X \otimes I_2^{\otimes p}\otimes I_d=\pm 1$}.}.
\end{lemma}
\begin{proof}
From Eq.~\eqref{eq:homomorphism_product} and $ZH=HX$, we have
\begin{equation}
    B(I_2^{\otimes k-p} \otimes Z \otimes I_2^{\otimes p}\otimes I_d)B(H^{\otimes k+1} \otimes I_d) = B(H^{\otimes k+1} \otimes I_d) B(I_2^{\otimes k-p} \otimes X \otimes I_2^{\otimes p}\otimes I_d).
\end{equation}
Therefore, the output state $\ket*{\Psi'_\pm}=B(H^{\otimes k+1} \otimes I_d)\ket*{\Psi_\pm}$ satisfies
\begin{equation}
    B(I_2^{\otimes k-p} \otimes Z \otimes I_2^{\otimes p} \otimes I_d) \ket*{\Psi'_\pm} = \pm \ket*{\Psi'_\pm}\label{eq:eigenstate}
\end{equation}
because of Eq.~\eqref{eq:symmetry_condition}.
Let $S_\pm$ be the set of the measurement patterns that span the eigenspace of $B(I_2^{\otimes k-p} \otimes Z \otimes I_2^{\otimes p} \otimes I_d)$ with the eigenvalue of $\pm 1$, that is, 
\begin{equation}
    S_\pm = \{\ket{\vb*{n}} \mid B(I_2^{\otimes k-p} \otimes Z \otimes I_2^{\otimes p} \otimes I_d)\ket{\vb*{n}}=\pm \ket{\vb*{n}}\}.
\end{equation}
The input state $\ket*{\Psi_\pm}$ does not give any measurement pattern $\ket*{\vb*{n}} \in S_{\mp}$ because
\begin{equation}
    \bra*{\vb*{n}}\ket*{\Psi'_\pm}= \pm \bra*{\vb*{n}}B(I_2^{\otimes k-p} \otimes Z \otimes I_2^{\otimes p} \otimes I_d)\ket*{\Psi'_\pm}=-\bra*{\vb*{n}}\ket*{\Psi'_\pm},
\end{equation}
that is, $\bra*{\vb*{n}}\ket*{\Psi_\pm} = 0$, where the first equality holds from Eq.~\eqref{eq:eigenstate}, and the second equality holds from $\ket*{\vb*{n}} \in S_{\mp}$.
Therefore, $\ket*{\Psi_+}$ and $\ket*{\Psi_-}$ are distinguished from whether the obtained measurement pattern $\ket*{\vb*{n}}$ belongs to $S_+$ or $S_-$, which can be written down concretely as
\begin{equation}\label{eq:lemma_condition}
    \ket{\vb*{n}} \in S_+ \Leftrightarrow \sum_{i=1}^{2^{k-p-1}}\sum_{j=0}^{2^pd-1}n_{(2i-1)2^p d+j}=0 \pmod{2},  \quad \ket{\vb*{n}} \in S_- \Leftrightarrow \sum_{i=1}^{2^{k-p-1}}\sum_{j=0}^{2^pd-1}n_{(2i-1)2^p d+j}=1 \pmod{2}.
\end{equation}
\end{proof}

\subsubsection{The case of a Bell-state ancilla}
First, let us consider the case of $k=1$, that is, the LOC consists of the interferometric part specified by $H^{\otimes 2} \otimes I_d$, four detectors, and ancilla state $\ket*{\phi^\nu}=\sqrt{d^{-1}}\sum_{i\in \mathbb{Z}_d}\ket{\nu_i}\ket{\nu_i}$, as shown in Fig.~\ref{fig:graphical_ancilla}.
As in the manuscript, we divide the measurement patterns $\{\ket*{\vb*{n}}\}$ into cases according to $N_i=\sum_{k=0}^3 a_{dk+i}^\dag a_{dk+i}$ for $i\in \mathbb{Z}_d = \{0,\dots, d-1\}$, which is the sum of the photon numbers detected on the modes in $S_i =\{j\in \{0,\cdots, 4d-1\} \mid j=i \pmod{d}\}$.
The interferometric part of the LOC represented by $H^{\otimes 2} \otimes I_d$ preserves the values of $(N_i)_{i\in \mathbb{Z}_d}$ because it consists of the $d$ disjoint LOCs, each of which is represented by $H^{\otimes 2}$ and acts only among the modes in $S_i$.
It means that the measured value of $N_i$ is equivalent to the total number of input photons incident on the modes in $S_i$.
While a photon of each input qudit contributes to increasing $N_i$ by one for $i \in \mathbb{Z}_d$, two photons of the ancilla state $\ket*{\phi^\nu}$ contribute to increasing $N_j$ by two for $j \in \mathbb{Z}_d$.
Therefore, the measurement patterns are classified to the four cases: (i) $N_i=4$, (ii) $N_i=3$ and $N_j=1$, (iii) $N_i=2$ and $N_j=2$, and (iv) $N_i=1$, $N_j=1$, and $N_k=2$ for different $i, j, k \in \mathbb{Z}_d$.

We begin by deriving the Kraus operators corresponding to a measurement pattern satisfying either (ii) $N_i=3$ and $N_j=1$ or (iv) $N_i=1$, $N_j=1$, and $N_k=2$ for different $i, j, k \in \mathbb{Z}_d$.
Let us consider an orthonormal basis including $\ket*{\psi_{i,j,+}}$ and $\ket*{\psi_{i,j,-}}$ as $\{\ket*{\psi_{i,j,+}},\ket*{\psi_{i,j,-}}, \cdots \}$ for the input two qudits.
The input states that can give such a measurement pattern are only $\ket*{\psi_{i,j,+}}$ or $\ket*{\psi_{i,j,-}}$ among the basis states, where the corresponding entire input state to the LOC is $\ket*{\psi^\nu_{i,j,\pm}}\ket*{\phi^\nu}$. 
In addition, these input states satisfy
\begin{equation}\label{eq:condition_psi}
    B(I_2 \otimes X \otimes I_d)\ket*{\psi^\nu_{i,j,\pm}}\ket*{\phi^\nu} = B((X \otimes I_d) \oplus (X \otimes I_d))\ket*{\psi^\nu_{i,j,\pm}}\ket*{\phi^\nu} = B(X \otimes I_d)\ket*{\psi^\nu_{i,j,\pm}}\otimes B(X \otimes I_d)\ket*{\phi^\nu} = \pm \ket*{\psi^\nu_{i,j,\pm}}\ket*{\phi^\nu},
\end{equation}
where the second equality holds from Eq.~\eqref{eq:homomorphism_tensor} and the third equality holds because $B(X \otimes I_d)$ corresponds to the exchange of the two qudits, and thereby $B(X \otimes I_d)\ket*{\psi^{\nu}_{i,j,\pm}}=\pm \ket*{\psi^{\nu}_{i,j,\pm}}$ and $B(X \otimes I_d) \ket*{\phi^{\nu}}= \ket*{\phi^{\nu}}$ hold.
From the Lemma for $k=1$ and $p=0$, Eq.~\eqref{eq:condition_psi} means that $\ket*{\psi^\nu_{i,j,+}}\ket*{\phi^\nu}$ and $\ket*{\psi^\nu_{i,j,-}}\ket*{\phi^\nu}$ are distinguished from the measurement pattern (particularly based on which condition in Eq.~\eqref{eq:lemma_condition} is satisfied).
In other words, when the obtained measurement pattern belongs to case (ii) or case (iv), we can determine whether the input state of two qudits is $\ket*{\psi_{i,j,+}}$ or $\ket*{\psi_{i,j,-}}$ from the measurement pattern, which corresponds to the application of the Kraus operators $\bra*{\psi_{i,j,+}}$ or $\bra*{\psi_{i,j,-}}$, respectively.

Next, we derive the Kraus operator corresponding to a measurement pattern satisfying (i) $N_i=4$ for $i \in \mathbb{Z}_d$.
For an orthonormal basis $\{\ket{i}\ket{i}, \cdots\}$ for the input two qudits, $\ket{i}\ket{i}$ is the unique input state that can give such a measurement pattern among the basis states.
The total probability of obtaining such measurement patterns is $d^{-1}$, which is computed from the corresponding entire input state to the LOC,
\begin{equation}\label{eq:one_ancilla_decomp1}
    \ket*{\nu_{i}}\ket*{\nu_{i}}\ket*{\phi^{\nu}}=\frac{1}{\sqrt{d}}\ket*{\nu_i}^{\otimes 2} \sum_{j \neq i}\ket*{\nu_j}^{\otimes 2} + \frac{1}{\sqrt{d}}\ket*{\nu_i}^{\otimes 4},
\end{equation}
where only the second term contributes to the case of $N_i=4$.
Therefore, the corresponding Kraus operator is $\sqrt{d^{-1}}\bra{i}\bra{i}$.

Finally, we derive the Kraus operators corresponding to a measurement patterns satisfying (ii) $N_i=2$ and $N_j=2$ for different $i, j \in \mathbb{Z}_d$. 
For an orthonormal basis of $\{\ket*{\phi_{i,j,+}},\ket*{\phi_{i,j,-}},\cdots
\}$ for the input two qudits, $\ket*{\phi_{i,j,+}}$ and $\ket*{\phi_{i,j,-}}$ are the unique input states that can give such a measurement pattern among the basis states, and the corresponding entire input states are decomposed as
\begin{align}\label{eq:one_ancilla_decomp2}
    \ket*{\phi^\nu_{i,j,\pm}}\ket*{\phi^{\nu}} = 
    \frac{1}{\sqrt{2d}}(\ket*{\nu_i}^{\otimes 2}\ket*{\nu_j}^{\otimes 2} \pm \ket*{\nu_j}^{\otimes 2}\ket*{\nu_i}^{\otimes 2}) 
    + \frac{1}{\sqrt{d}}\ket*{\phi^\nu_{i,j,\pm}}\sum_{k \neq i,j}\ket*{\nu_k}^{\otimes2}
    + \frac{1}{\sqrt{2d}}(\ket*{\nu_i}^{\otimes 4}\pm\ket*{\nu_j}^{\otimes 4}).
\end{align}
The first term of $\sqrt{2^{-1}}(\ket*{\nu_i}^{\otimes 2}\ket*{\nu_j}^{\otimes 2} \pm \ket*{\nu_j}^{\otimes 2}\ket*{\nu_i}^{\otimes 2})$ with the coefficient $\sqrt{d^{-1}}$ corresponds to the case of $N_i=2$ and $N_j=2$.
The first term satisfies that
\begin{equation}
    B(X \otimes I_2 \otimes I_d)\frac{1}{\sqrt{2}}(\ket*{\nu_i}^{\otimes 2}\ket*{\nu_j}^{\otimes 2} \pm \ket*{\nu_j}^{\otimes 2}\ket*{\nu_i}^{\otimes 2}) = \pm \frac{1}{\sqrt{2}}(\ket*{\nu_i}^{\otimes 2}\ket*{\nu_j}^{\otimes 2} \pm \ket*{\nu_j}^{\otimes 2}\ket*{\nu_i}^{\otimes 2})
\end{equation}
because $B(X \otimes I_2 \otimes I_d)$ corresponds to the exchange of first two qudits and last two qudits.
Thus, the Lemma can be applied for $k=1$ and $p=1$, which concludes that these two states $\sqrt{2^{-1}}(\ket*{\nu_i}^{\otimes 2}\ket*{\nu_j}^{\otimes 2} \pm \ket*{\nu_j}^{\otimes 2}\ket*{\nu_i}^{\otimes 2})$ are distinguished from the measurement pattern.
Therefore, for input states $\ket*{\phi_{i,j,+}}$ or $\ket*{\phi_{i,j,+}}$ of the input two qudits, a measurement pattern satisfying case (ii) is obtained with the probability of $d^{-1}$, and then the input state is determined whether $\ket*{\phi_{i,j,+}}$ or $\ket*{\phi_{i,j,+}}$ from the measurement pattern, which means that the corresponding Kraus operators are $\sqrt{d^{-1}}\bra*{\phi_{i,j,+}}$ or $\sqrt{d^{-1}}\bra*{\phi_{i,j,-}}$.

At the end, we have completed the identification of the Kraus operators for all measurement patterns as 
\begin{equation}\label{eq:one_ancilla}
    \{\sqrt{d^{-1}}\bra{i}\bra{i}, \sqrt{d^{-1}}\bra*{\phi_{i,j,+}}, \sqrt{d^{-1}}\bra*{\phi_{i,j,-}}, \bra*{\psi_{i,j,+}}, \bra*{\psi_{i,j,+}} \mid i,j \in \mathbb{Z}_d, i<j\}.
\end{equation}

\subsubsection{The general cases}
Here, we prove the case of any integer $k \geq 0$.
The LOC consists of the interferometric part specified by $H^{\otimes k+1} \otimes I_d$, $2^{k+1} d$ detectors, and $2(2^k-1)$ ancilla photons in the state $\bigotimes_{q=1}^{k} \ket*{\text{GHZ}^{\nu}(2^q)}$, where
$\ket*{\text{GHZ}(2^q)} = \sqrt{d^{-1}} \sum_{i\in \mathbb{Z}_d} \ket{i}^{\otimes 2^q}$ is a $2^q$-qudit GHZ state.
The LOC for $k=2$ is shown in Fig.~\ref{fig:second_order}.
The interferometric part of the LOC represented by transfer matrix $H^{\otimes k+1} \otimes I_d$ preserves the value of $N_i=\sum_{k=0}^{2^{k+1}-1} a_{dk+i}^\dag a_{dk+i}$ for any $i\in \mathbb{Z}_d$ because it consists of $d$ disjoint LOCs, each of which is represented by $H^{\otimes k+1}$ and acts among the modes in $S_i =\{j\in \{0,\cdots, 2^{k+1}d-1\} \mid j=i \pmod{d}\}$.
Thus, we can classify the obtained measurement patterns $\{\ket*{\vb*{n}}\}$ according to $\vb*{N} \equiv (N_i)_{i \in \mathbb{Z}_d}$.
For a state $\ket{i}\ket{i}$ of the input two qudits for $i \in \mathbb{Z}_d$, let us consider the following decomposition of the corresponding entire input state:
\begin{equation}\label{eq:state_decomposition}
    \ket{\nu_i}\ket{\nu_i}\bigotimes_{q=1}^k\ket*{\text{GHZ}^\nu(2^q)}
    = \sum_{p=1}^{k-1} \qty\Big( \frac{1}{\sqrt{d^p}}\ket{\nu_i}^{\otimes 2^{p}}\sum_{j\neq i} \ket{\nu_j}^{\otimes 2^{p}} \bigotimes_{q=p+1}^k\ket*{\text{GHZ}^\nu(2^q)} ) + \frac{1}{\sqrt{d^k}}\ket{\nu_i}^{\otimes 2^{k}}\sum_{j\neq i} \ket{\nu_j}^{\otimes 2^{k}} + \frac{1}{\sqrt{d^{k}}}\ket{\nu_i}^{\otimes 2^{k+1}},
\end{equation}
which is obtained by recursively applying the transformation of
\begin{equation}
    \ket*{\nu_i}^{\otimes 2^p}\bigotimes_{q=p}^k \ket*{\text{GHZ}^\nu(2^q)} =  \frac{1}{\sqrt{d}}\ket*{\nu_i}^{\otimes 2^{p}} \sum_{j\neq i} \ket{\nu_j}^{\otimes 2^{p}} \bigotimes_{q=p+1}^k\ket*{\text{GHZ}^\nu(2^q)}+ \frac{1}{\sqrt{d}}\ket*{\nu_i}^{\otimes 2^{p+1}}\bigotimes_{q=p+1}^k\ket*{\text{GHZ}^\nu(2^q)}
\end{equation}
from $p=1$ to $p=k$. 
On the $p$th term in the right-hand side of Eq.~\eqref{eq:state_decomposition} for $p=1,\dots, k$, the first and second $2^p$ qudits are in the state $\ket*{\nu_i}$ and $\ket*{\nu_j}$ for $j (\neq i) \in \mathbb{Z}_d$, respectively, and contribute to increasing $N_i$ and $N_j$ by $2^{p}$.
Except for $p=k$, the other qudits are in the GHZ states whose sizes are at least $2^{p+1}$. Thus, they contribute to increasing $N_k$ by at least $2^{p+1}$ for $k \in \mathbb{Z}$ (which may be the same as $i$ or $j$).
Therefore, the value of $p$ can be computed from $\vb*{N}$ as $v(\vb*{N})=\min_{i=0}^{d-1}f(N_i)-1$, where $f(N_i) =\min \{q\in \{1,\dots,k+2\} \mid 2^{q} \nmid N_i\}$ is the smallest digit having a value of $1$ when $N_i$ is represented in binary.
In addition, we define $I(\vb*{N})=\argmin_{i=0}^{d-1} f(N_i)$.
Then, the $p$th term in Eq.~\eqref{eq:state_decomposition} corresponds to the case of $v(\vb*{N})=p$ for $p=1,\dots, k+1$, where $I(\vb*{N})=\{i,j\}$ for $p \neq k+1$ and $I(\vb*{N})=\{i\}$ for $p = k+1$.
For a state $\ket{i}\ket{j}$ of the input two qudits for different $i, j \in \mathbb{Z}_d$, the entire input state $\ket{\nu_i}\ket{\nu_j}\bigotimes_{q=1}^k\ket*{\text{GHZ}^\nu(2^q)}$ always gives $v(\vb*{N})=0$ because of the contribution from $\ket{\nu_i}\ket{\nu_j}$, where $I(\vb*{N})=\{i,j\}$.
Therefore, we can classify the measurement patterns according to $v(\vb*{N})$ and $I(\vb*{N})$, instead of $\vb*{N}$ itself, which greatly reduces the number of cases we need to consider.

We begin by the case of $v(\vb*{N})=0$ and $I(\vb*{N})=\{i,j\}$. 
For an orthonormal basis of $\{\ket*{\psi_{i,j,+}},\ket*{\psi_{i,j,-}}, \cdots \}$ for the input two qudits, $\ket*{\psi_{i,j,+}}$ and $\ket*{\psi_{i,j,-}}$ are the unique input states that can give $\vb*{N}$ satisfying $v(\vb*{N})=0$ and $I(\vb*{N})=\{i,j\}$ among the basis states.
The entire input states corresponding to them satisfy
\begin{align}
    B(I_2^{\otimes k} \otimes X \otimes I_d)\ket*{\psi^{\nu}_{i,j,\pm}}\bigotimes_{q=1}^k\ket*{\text{GHZ}^{\nu}(2^q)} &= (B(X \otimes I_d)\ket*{\psi^{\nu}_{i,j,\pm}})  \bigotimes_{q=1}^k (B(I_2^{\otimes q-1} \otimes X \otimes I_d) \ket*{\text{GHZ}^{\nu}(2^q)}) \notag\\
    &=\pm \ket*{\psi^{\nu}_{i,j,\pm}}\bigotimes_{q=1}^k\ket*{\text{GHZ}^{\nu}(2^q)},
\end{align}
where the first equality holds from Eq.~\eqref{eq:homomorphism_tensor} and the second equality holds from the invariance of the GHZ states under arbitrary permutation of the qudits.
Therefore, these states are distinguished from the measurement pattern from the Lemma for $p=0$.
Thus, this case corresponds to the application of the Kraus operator $\bra*{\psi_{i,j,+}}$ or $\bra*{\psi_{i,j,-}}$.

For the case of $v(\vb*{N})=k+1$ and $I(\vb*{N})=\{i\}$, we consider an orthonormal basis $\{\ket{i}\ket{i}, \cdots\}$ for the input two qudits.
Then, $\ket{i}\ket{i}$ is the unique input state that can give $\vb*{N}$ satisfying $v(\vb*{N})=k+1$ and $I(\vb*{N})=\{i\}$ among the basis states.
Thus, the corresponding Kraus operator is determined to be $\sqrt{d^{-k}}\bra{i}\bra{i}$, where the coefficient $\sqrt{d^{-k}}$ is derived from Eq.~\eqref{eq:state_decomposition}.

Finally, we identify the Kraus operators corresponding to the case of $v(\vb*{N}) \neq 0, k+1$ and $I(\vb*{N})=\{i,j\}$.
We consider an orthonormal basis of $\{\ket*{\phi_{i,j,+}},\ket*{\phi_{i,j,-}},\dots\}$ for the input two qudits.
Then, $\ket*{\phi_{i,j,+}}$ and $\ket*{\phi_{i,j,-}}$ are the unique input states that can give $\vb*{N}$ satisfying $v(\vb*{N}) \neq 0, k+1$ and $I(\vb*{N})=\{i,j\}$ among the basis states.
The entire input states are decomposed from Eq.~\eqref{eq:state_decomposition} as
\begin{align}\label{eq:state_decomposition2}
    \ket{\phi^\nu_{i,j,\pm}}\bigotimes_{q=1}^k\ket*{\text{GHZ}^\nu(2^q)}
    &= \sum_{p=1}^{k-1}\qty\Big( \frac{1}{\sqrt{d^{p}}} \ket*{\Xi^\nu_{i,j,\pm}(2^{p+1})} \bigotimes_{q=p+1}^k\ket*{\text{GHZ}^\nu(2^q)}) + \frac{1}{\sqrt{d^{k}}} \ket*{\Xi^\nu_{i,j,\pm}(2^{k+1})} \notag \\
    &+\sum_{p=1}^{k-1} \qty\Big(\frac{1}{\sqrt{d^{p}}}\ket*{\text{GHZ}^\nu_{i,j,\pm}(2^{p})}\sum_{l\neq i,j} \ket*{\nu_l}^{\otimes 2^{p}}\bigotimes_{q=p+1}^k\ket*{\text{GHZ}^\nu_{i,j,\pm}(2^q)}) \notag \\
    &+\frac{1}{\sqrt{d^{k}}}\ket*{\text{GHZ}^\nu_{i,j,\pm}(2^{k})}\sum_{l\neq i,j} \ket*{\nu_l}^{\otimes 2^{k}}
    + \frac{1}{\sqrt{d^{k}}}\ket*{\text{GHZ}^\nu_{i,j,\pm}(2^{k+1})},
\end{align}
where
\begin{align}
    \ket*{\Xi_{i,j,\pm}(2n)} &= \frac{1}{\sqrt{2}}(\ket{i}^{\otimes n}\ket{j}^{\otimes n} \pm \ket{j}^{\otimes n}\ket{i}^{\otimes n}), \\
    \ket*{\text{GHZ}_{i,j,\pm} (n)} &= \frac{1}{\sqrt{2}}(\ket{i}^{\otimes n} \pm \ket{j}^{\otimes n}).
\end{align}
Only the first line in the right-hand side of Eq.~\eqref{eq:state_decomposition2} is relevant to the case of $v(\vb*{N}) \neq 0, k+1$ and $I(\vb*{N})=\{i,j\}$, where the $p$th term corresponds to the case of $v(\vb*{N}) = p$ for $p=1,\dots, k$.
The $k$th term satisfies
\begin{equation}
    B( X \otimes I_2^{\otimes k} \otimes I_d) \ket*{\Xi^{\nu}_{i,j,\pm}(2^{k+1})} = \pm \ket*{\Xi^{\nu}_{i,j,\pm}(2^{k+1})},
\end{equation}
and the other $p$th term satisfies
\begin{align}
    &B(I_2^{\otimes k-p} \otimes X \otimes I_2^{\otimes p} \otimes I_d) \ket*{\Xi^{\nu}_{i,j,\pm}(2^{p+1})}\bigotimes_{q=p+1}^k\ket*{\text{GHZ}^{\nu}(2^q)} \notag  \\ &= (B(X \otimes I_2^{\otimes p} \otimes I_d) \ket*{\Xi^{\nu}_{i,j,\pm}(2^{p+1})})\bigotimes_{q=p+1}^k B(I_2^{\otimes q-p-1} \otimes X \otimes I_2^{\otimes p} \otimes I_d) \ket*{\text{GHZ}^{\nu}(2^q)} = \pm \ket*{\Xi^{\nu}_{i,j,\pm}(2^{p+1})}\bigotimes_{q=p+1}^k\ket*{\text{GHZ}^{\nu}(2^q)}
\end{align}
because $B(X \otimes I_2^{\otimes p} \otimes I_d)$ corresponds to the exchange of the first $2^p$ qudits and the last $2^p$ qudits, and thereby $B(X \otimes I_2^{\otimes p} \otimes I_d)\ket*{\Xi_{i,j,\pm}(2^{p+1})} = \pm \ket*{\Xi_{i,j,\pm}(2^{p+1})}$ holds.
From the Lemma for $p=1, \dots, k$, these two states are discriminated from the measurement pattern.
Therefore, this case corresponds to the application of the Kraus operator $\sqrt{{c_k}}\bra*{\phi_{i,j,\pm}}$, where the coefficient $c_k$ is determined to be $c_k=\sum_{p=1}^k d^{-p}$ by summing up the coefficient $d^{-p}$ of the $p$th term in Eq.~\eqref{eq:state_decomposition2} for $p=1,\cdots,k$.

From these results, we have identified the quantum measurement to which the considered LOC corresponds as
\begin{equation}
    \{\sqrt{d^{-k}}\bra{i}\bra{i}, \sqrt{c_k}\bra*{\phi_{i,j,+}}, \sqrt{c_k}\bra*{\phi_{i,j,-}}, \bra*{\psi_{i,j,+}}, \bra*{\psi_{i,j,-}} \mid i,j\in \mathbb{Z}_d,i < j\}.
\end{equation}
We can ensure that other Kraus operators are not included in the quantum measurement from the fact that the sum of its POVM elements is equal to the identity operator as
\begin{equation}
    \frac{1}{d^k} \sum_i \ket{i}\ket{i}\bra{i}\bra{i} +  \frac{1}{c_k}\sum_{i < j} (\ket*{\phi_{i,j,+}}\bra*{\phi_{i,j,+}}+\ket*{\phi_{i,j,-}}\bra*{\phi_{i,j,-}}) + \sum_{i < j}(\ket*{\psi_{i,j,+}}\bra*{\psi_{i,j,+}}+\ket*{\psi_{i,j,-}}\bra*{\psi_{i,j,-}})=\sum_{i,j}\ket{i}\ket{j}\bra{i}\bra{j},
\end{equation}
where $d^{-k}+(d-1)c_k^{-1}=1$ is used.

\subsection{B. Performance comparison among quantum repeater protocols}
Here, we evaluate the performance of the quantum repeater protocols we proposed in the manuscript and compare them with standard quantum repeater protocols using linear-optical fusion gates on qubits.
The first- and second-generation quantum repeater protocols commonly use a heralded entanglement generation scheme to produce entanglement links between adjacent nodes while they are different in how to perform the entanglement swapping.
The central idea of our protocol is to replace the standard fusion gates with the pairwise fusion gates and the standard entanglement swapping with the boosted entanglement swapping.
Therefore, we describe the first- and second-generation quantum repeater protocols independently of which fusion gates and entanglement swapping are used.
For simplicity, we assume that the local generation of initial states is deterministic and fast enough to be able to neglect its generation time, where the initial states means Bell states on photonic qubits when the standard fusion gates are to be used and three-partite GHZ states on photonic qudits when the pairwise fusion gates are to be used in the protocol.
In addition, we ignore any error other than photon loss and the probabilistic nature of the linear-optical fusion gates.

When we take the detection efficiency $\eta_\text{d}$ of detectors into account, the success probabilities of the fusion gates are represented as $\eta_\text{d}^2 p_\text{f}$, where $p_\text{f}$ is the intrinsic success probability of each linear-optical fusion gate, that is, $p_\text{f}=1/2$ for the standard fusion gate and $p_\text{f}=1-d^{-1}$ for the pairwise fusion gate without ancilla photons (and $p_\text{f}=(1-d^{-(k+1)})\eta_\text{d}^{2(2^k-1)}$ for the pairwise fusion gates with $2(2^k-1)$ ancilla photons).
Let $P_\text{s}$ be the success probability of each entanglement swapping and $\eta_\text{c}$ be the coupling efficiency between a photon and a quantum memory. Then, $P_\text{s}=\eta_\text{d}^2 \eta_\text{c}^2 p_\text{f}$ for the standard entanglement swapping, and $P_\text{s}=\eta_\text{d}^4 \eta_\text{c}^4 p_\text{f}$ for the boosted entanglement swapping, where the additional factor $\eta_\text{d}^2 \eta_\text{c}^2$ is caused by the use of additional photons in the GHZ states.

\subsubsection{Entanglement generation step}
In the entanglement generation step, each repeater node sends a photon of a locally-prepared initial state to each adjacent intermediate node at a distance of $L_0/2$ and preserves the other photons in local quantum memories.
Each intermediate node performs a (standard or pairwise) fusion gate on the received two photons.
The transmission rate of an optical fiber of length $L$ is written as $\eta_\text{t}(L)=e^{-L/L_\text{att}}$, where $L_\text{att}=\SI{22}{\km}$ is a typical attenuation length of telecom-wavelength optical fibers.
Then, the success probability of this entanglement generation process is $P_\text{g}(L_0)=\eta_\text{t}(L_0) \eta_\text{d}^2 p_\text{f}$.
Each repeater node repeats this process until success.
The cycle time of this trial is $\tau_0 = L_0/c'$, where $c'= \SI{2e5}{\km\per\s}$ is a typical speed of light in telecom-wavelength optical fibers, because both of the transmissions of the photon and the heralding signal to announce the result of the entanglement generation take $\tau_0/2$.
The average time for each entanglement generation to succeed is 
\begin{equation}\label{eq:T0}
    T_0=\frac{\tau_0}{P_\text{g}(L_0)}=\frac{\tau_0}{\eta_\text{d}^2 p_\text{f}}e^{L/L_\text{att}}.
\end{equation}

\subsubsection{Entanglement swapping step in first-generation quantum repeaters}
In first-generation quantum repeaters, each entanglement swapping is also implemented in a heralded manner.
We consider an $n$-level nested structure, where $2^n -1$ repeater nodes are equally spaced between Alice and Bob, and the entire distance between them is $L_\text{tot} = 2^n L_0$.
Let us call each repeater node from Alice to Bob the $i$th repeater node for $i=1,\dots, 2^n-1$.
Then, once the $(2i-1)$th repeater node for $i=1,\dots, 2^{n-1}$ confirms the success of the both sides of the entanglement generation with the adjacent repeater nodes at a distance of $L_0$, it retrieves a pair of the photons from the quantum memories, performs the entanglement swapping on the pair, and sends the heralding signal to the adjacent nodes. 
We call this process the first level of entanglement swapping.
Similarly, once the $(2^ki-2^{k-1})$th repeater node for $i=1,\dots, 2^{n-k}$ confirms the success of the both sides of the entanglement swappings in the repeater nodes at a distance of $2^{k-1} L_0$ from the heralding signals, it retrieves a pair of the photons from the quantum memories, performs the entanglement swapping on the pair, and sends the heralding signal to the nodes at a distance of $2^{k} L_0$.
We call this process the $k$th level of entanglement swapping.
Let $\tau_k$ be the average cycle time from the beginning of the entanglement generation step to the implementation of the $k$th level of entanglement swapping.
Then, the average time for the $k$th level of entanglement swapping to succeed is $T_k=\tau_k/{P_\text{s}}$ for $k\geq 1$, and it holds that
\begin{equation}\label{eq:recursive}
    \tau_{k+1} \simeq (3/2)T_{k} + 2^{k}\tau_0
\end{equation}
for $k=0,1,\dots,n-1$.
The factor of $3/2$ comes from the fact that each repeater node has to wait for the both sides of the entanglement links to be ready~\cite{Sangouard_Rev.Mod.Phys.2011}.
The second term is the transmission time of the heralding signal to inform the nodes at a distance of $2^{k}L_0$ of the result of the $k$th level of the entanglement swapping~\footnote{This heralding signal is necessary even when the $k$th level of entanglement swapping failed, because the $(2^{k+1}i-2^k)$th repeater node cannot discard the photons preserved in the quantum memories and restart the entanglement generation step until it knows that the $k$th level of entanglement swapping failed}.
Eq.~\eqref{eq:recursive} is solved with Eq.~\eqref{eq:T0} as 
\begin{equation}
    T_k = 2^k \tau_0\qty(\frac{\alpha^k}{P_\text{g}(L_0)}+\frac{2\alpha(\alpha^{k}-1)}{3(\alpha-1)}),
\end{equation}
where $\alpha = 3/(4P_\text{s})$.
Although the average time to generate a Bell state between Alice and Bob is $T_n$, each repeater node, including Alice and Bob, can repeat the process without waiting for the heralding signal of the last level of entanglement swapping in many cases, such as Alice and Bob have a sufficient number of quantum memories or immediately measure the shared state for quantum key distribution~\footnote[10]{The heralding signal to announce the result of the last entanglement swapping is also necessary for the synchronization among all the repeater node. In principle, such synchronization is unnecessary because each repeater node can start the next entanglement generation step immediately after performing the entanglement swapping although the entanglement generation will never succeed until the adjacent node also finishes the entanglement swapping and starts the entanglement generation step. However, the absence of the synchronization causes a difference in the timing for the entanglement generation step to start among the repeater nodes, which in turn can make, for example, the required memory time longer. Here, for simplicity, we ignore such effects although the transmission time of the heralding signal for the synchronization is not included in the cycle time of the protocol}.
Thus, we let the average distribution time per entangled pair in the first-generation protocol be $T^*_\text{first}$ defined as
\begin{equation}\label{eq:first_generation}
    T_\text{first}(n)=\frac{3}{2P_\text{s}}T_{n-1} = \frac{L_\text{tot}}{c'}\qty(\frac{\alpha^n}{P_\text{g}(L_\text{tot}/2^n)}+\frac{2\alpha^2(\alpha^{n-1}-1)}{3(\alpha-1)}), \quad T^*_\text{first} = \min_{n=1,2,\dots} T_\text{first}(n),
\end{equation}
where the number of repeater nodes is optimized to minimize the distribution time.
The memory time of the quantum memories required for the optimized protocol to operate is $\tilde{T}_\text{first}=(3/2)T_{n^*-1}$ for $n^* = \argmin_n T_\text{first}(n)$, that is, $P_s T^*_\text{first}$.

Interestingly, the scaling of $T^*_\text{first}$ with respect to $L_\text{tot}$ changes depending on whether $\alpha < 1$ or $\alpha > 1$, where $\alpha < 1$ corresponds to the case of the success probability of the entanglement swapping higher than $3/4$, which is not achievable in the conventional protocol using the standard entanglement swapping.
In the case of $\alpha < 1$, we have $c'T_\text{first}(n)/L_\text{tot} \rightarrow 2\alpha^2/3(1-\alpha) = \text{const.}$ for $n \rightarrow \infty$.
Therefore, $T^*_\text{first}$ is also upper bounded by it, which implies that $T^*_\text{first}$ becomes linear with respect to $L_\text{tot}$~\footnote{$c'T_\text{first}(n)/L_\text{tot}$ is lower bounded by $2\alpha^2/3$ for $n>1$ and $\alpha/P_\text{g}(L_\text{tot}/2)$ for $n=1$. Thus, it is not only upper bounded but also lower bounded by a constant with respect to $L_\text{tot}$.}, that is,
\begin{equation}\label{eq:constant}
    \frac{c'}{L_\text{tot}}T^*_\text{first}(L_\text{tot}) = \text{const.}, \quad (\alpha < 1).
\end{equation}
On the other hand, in the case of $\alpha > 1$, let us consider to choose the number of repeater nodes as $n=\lceil \log_2(L_\text{tot}/L_\text{u}) \rceil$ for a constant $L_\text{u}$, instead of optimizing $n$. Then, we have
\begin{equation}
    T_\text{first}(n) < \frac{L_\text{tot}}{c'}\qty(\frac{\alpha(L_\text{tot}/L_\text{u})^{\log_2{\alpha}}}{P_\text{g}(L_\text{u})}+\frac{2\alpha^2((L_\text{tot}/L_\text{u})^{\log_2{\alpha}}-1)}{3(\alpha-1)})
\end{equation}
from $L_\text{tot}/2^{n}=L_0\leq L_\text{u}$ and $n<\log_2(L_\text{tot}/L_\text{u})+1$.
Therefore, $T^*_\text{first}$ is also upper bounded by it, which corresponds to
\begin{equation}\label{eq:upperbound}
    \frac{c'}{L_\text{tot}}T^*_\text{first}(L_\text{tot}) = O(L_\text{tot}^{\log_2{\alpha}}),  \quad (\alpha > 1).
\end{equation}
Furthermore, we can also prove that the order of $T^*_\text{first}$ is lower bounded as
\begin{equation}\label{eq:lowerbound}
    \frac{c'}{L_\text{tot}}T^*_\text{first}(L_\text{tot})=\omega\qty(\qty(\frac{L_\text{tot}}{\log{L_\text{tot}}})^{\log_2\alpha}),  \quad (\alpha > 1),
\end{equation}
where the symbol $\omega$ means that $f(x)/g(x) \rightarrow 0$ holds for $x\rightarrow \infty$ if we write $g(x) = \omega(f(x))$.
Therefore, the linear scaling of $T^*_\text{first}$ appears only when the success probability of entanglement swapping is higher than $3/4$; otherwise, the scaling of $T^*_\text{first}$ becomes $L_\text{tot}^{\log_2{\alpha}}$, except for the factor of $\log{L_\text{tot}}$.

\begin{proof}[Proof of Eq.~\eqref{eq:lowerbound}]
Let $\beta=e^{1/L_\text{att}} >1$ and $n^*(L_\text{tot}) = \argmin_n T_\text{first}(n)$.
In Eq.~\eqref{eq:first_generation}, the second term of $T_\text{first}(n)$ is positive and $P_\text{g}(L_\text{tot}/2^n)^{-1}\geq \beta^{L_\text{tot}/2^{n}}$ from $\eta_d^{2}p_\text{f}\leq 1$. Thus, we have $c'T^*_\text{first}/L_\text{tot}>\alpha^{n^*} \beta^{L_\text{tot}/2^{n^*}}$.
We derive a contradiction under the assumption of $c'T^*_\text{first}/L_\text{tot}=O(f(L_\text{tot}))$, where $f(L_\text{tot}) = (L_\text{tot}/\log_\beta{L_\text{tot}})^{\log_2\alpha}$.
We have $\alpha^{n^*} \beta^{L_\text{tot}/2^{n^*}} = O(f(L_\text{tot}))$ from this assumption.
In addition, since $\alpha^{n^*}$ and $\beta^{L_\text{tot}/2^{n^*}}$ are lower bounded by constants, respectively, we have $\beta^{L_\text{tot}/2^{n^*}}=O(f(L_\text{tot}))$ and $\alpha^{n^*} = O(f(L_\text{tot}))$.
These are equivalent to that, for large $L_\text{tot}$, there are constants $M$ and $M'$ such that $\beta^{L_\text{tot}/2^{n^*}} < M f(L_\text{tot})$ and $\alpha^{n^*} < M' f(L_\text{tot})$, respectively.
These inequalities lead to a lower bound and a upper bound of $n^*$ as
\begin{equation}
    n^* > n_\text{l} \equiv \log_2\qty(\frac{L_\text{tot}}{\log_\beta Mf(L_\text{tot})})
\end{equation}
and 
\begin{equation}
    n^* < n_\text{u} \equiv \log_\alpha M'f(L_\text{tot}),
\end{equation}
respectively. Therefore, we have
\begin{equation}\label{eq:inequality}
    \frac{\alpha^{n^*} \beta^{L_\text{tot}/2^{n^*}}}{f(L_\text{tot})} >\frac{\alpha^{n_\text{l}} \beta^{L_\text{tot}/2^{n_\text{u}}}}{f(L_\text{tot})} = \qty(\frac{L_\text{tot}^{(\log_\alpha{2})M'^{-\log_\alpha2}}\log_\beta L_\text{tot}}{\log_\beta M + (\log_2{\alpha})(\log_\beta L_\text{tot}-\log_\beta \log_\beta L_\text{tot})})^{\log_2 \alpha}.
\end{equation}
The right-hand side of Eq.~\eqref{eq:inequality} diverges for $L_\text{tot} \rightarrow \infty$, which is a contradiction.
\end{proof}

\subsubsection{Entanglement swapping step in second-generation quantum repeaters}
In second-generation quantum repeaters, every entanglement swapping is performed without waiting for heralding signals about other entanglement swappings.
Here, $m$ repeater nodes are equally spaced between Alice and Bob, where $L_\text{tot}=(m+1)L_0$.
Each repeater node implements the entanglement swapping once the both sides of the entanglement links are ready.
The average time for all the repeater nodes to complete the entanglement swapping is $\tau'_m \simeq H(m+1)T_0$, where the factor of $H(m+1)=\sum_{k=1}^{m+1}k^{-1}$ comes from the fact that all of the $m+1$ entanglement links should be established to complete all the entanglement swappings~\cite{Azuma_AVSQuantumSci.2021a}.
On the other hand, the heralding signal to inform Alice and Bob of the result of each entanglement swapping takes at most $m \tau_0$.
Thus, since the success probability of the entire entanglement swapping step is $P_\text{s}^m$, the average time for the generation of a Bell state between Alice and Bob is 
$(\tau'_m + m\tau_0)/P_\text{s}^m$.
However, as in the first-generation protocol, each repeater node, including Alice and Bob, can repeat the process without waiting for the heralding signals in many cases~\cite{Note10}.
Therefore, we let the average distribution time per entangled pair in the second-generation protocol be $T^*_\text{second}$ defined as 
\begin{equation}
    T_\text{second}(m) = \frac{\tau'_m}{P_\text{s}^m} = \frac{L_\text{tot}}{c'}\frac{H(m+1)}{(m+1)P_\text{s}^mP_\text{g}(L_\text{tot}/(m+1))}, \quad  T^*_\text{second} = \min_{m=1,2,\dots,} T_\text{second}(m),
\end{equation}
where the number of repeater nodes is optimized to minimize the distribution time.
The required coherence time of the quantum memories for the optimized protocol to operate is $\tilde{T}_\text{second} = \tau'_{m^*}$ for $m^* =\argmin_{m=1,2,\dots,} T_\text{second}(m)$.

Normally, $P_\text{s}$ is assumed to be unity in second-generation quantum repeaters.
In this case, by letting $m+1=\lceil L_\text{tot}/L_\text{u} \rceil $ for a constant $L_\text{u}$, it holds that 
\begin{equation}
    T_\text{second}(m) \leq \frac{L_\text{u}}{c'}\frac{H(\lceil L_\text{tot}/L_\text{u} \rceil)}{P_\text{g}(L_\text{u})}
\end{equation}
from $L_\text{tot}/(m+1) = L_0 \leq L_\text{u}$.
Therefore, we have
\begin{equation}
    T^*_\text{second}(L_\text{tot}) = O(\log{L_\text{tot}})
\end{equation}
from $H(n)=O(\log n)$.
This scaling of $T^*_\text{second}$ with respect to $L_\text{tot}$ is better than one of $T^*_\text{first}$.

\subsubsection{Numerical simulations}
\begin{figure}[tpb]
    \begin{minipage}[t]{0.48\linewidth}
    \centering
    \subcaption{}\label{fig:distribution_0.95}
    \hspace{-20pt}
    \includegraphics[height=5.5cm, trim=0 0 0 0, clip]{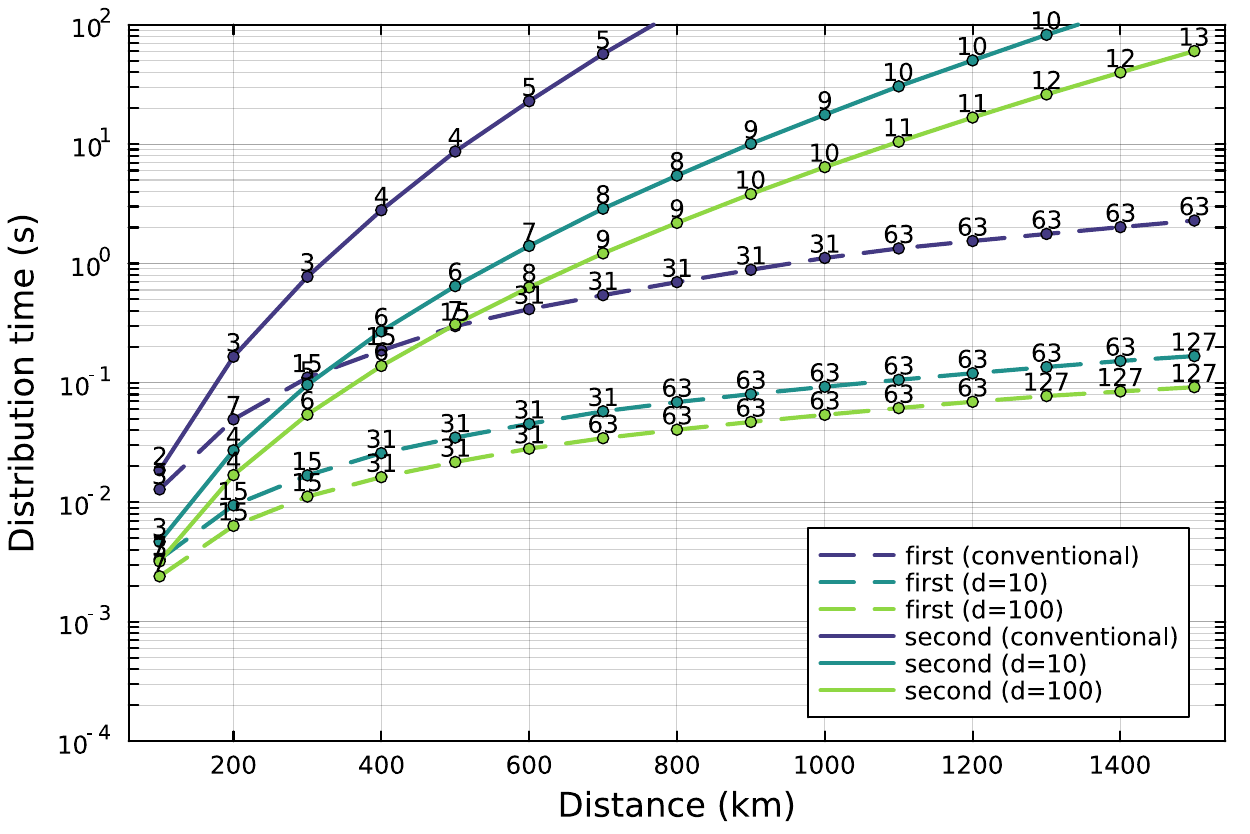}
    \vspace{-10pt}
    \end{minipage}
    \begin{minipage}[t]{0.48\linewidth}
    \centering
    \subcaption{}\label{fig:distribution_0.99}
    \hspace{-10pt}
    \includegraphics[height=5.5cm, trim=0 0 0 0, clip]{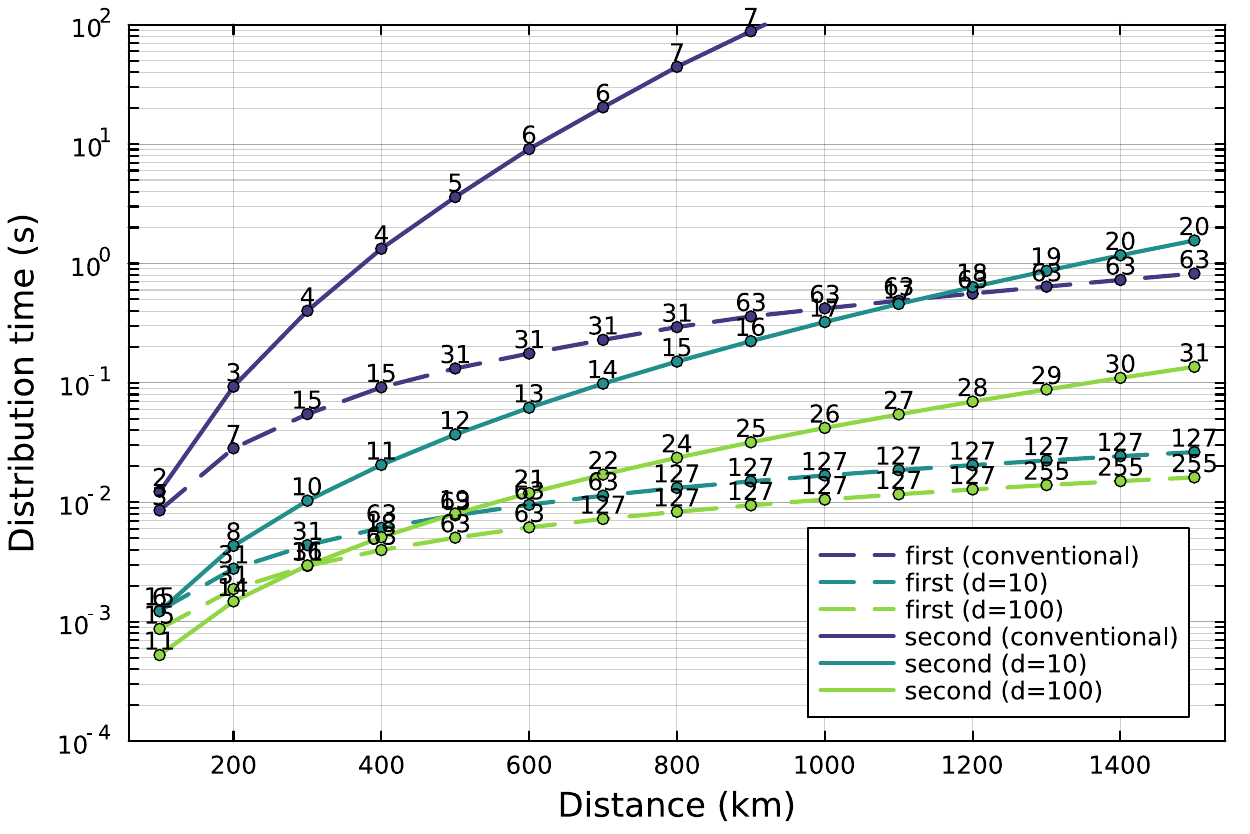}
    \end{minipage}
    \begin{minipage}[t]{0.48\linewidth}
        \centering
        \subcaption{}\label{fig:memory_0.99}
        \hspace{-20pt}
        \includegraphics[height=5.5cm, trim=0 0 0 0, clip]{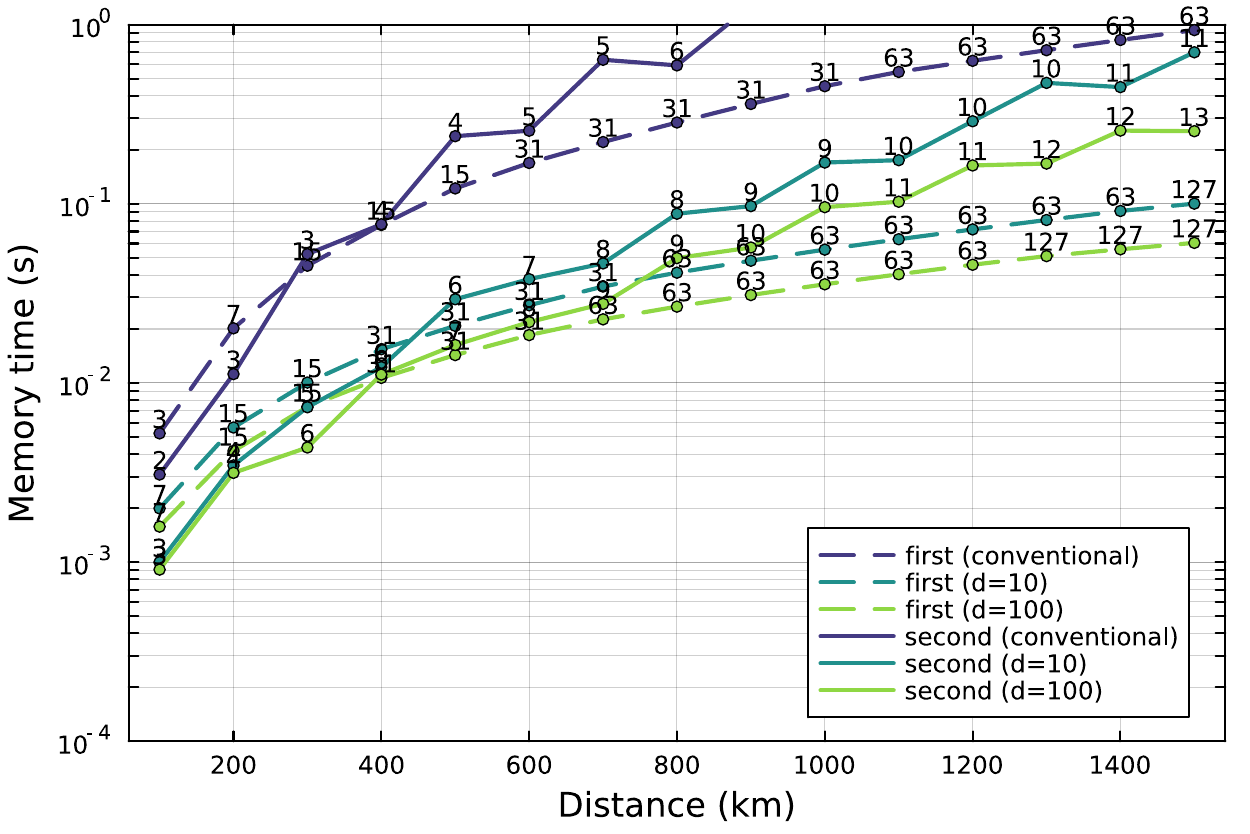}
        \vspace{-10pt}
        \end{minipage}
        \begin{minipage}[t]{0.48\linewidth}
        \centering
        \subcaption{}\label{fig:memory_0.99}
        \hspace{-10pt}
        \includegraphics[height=5.5cm, trim=0 0 0 0, clip]{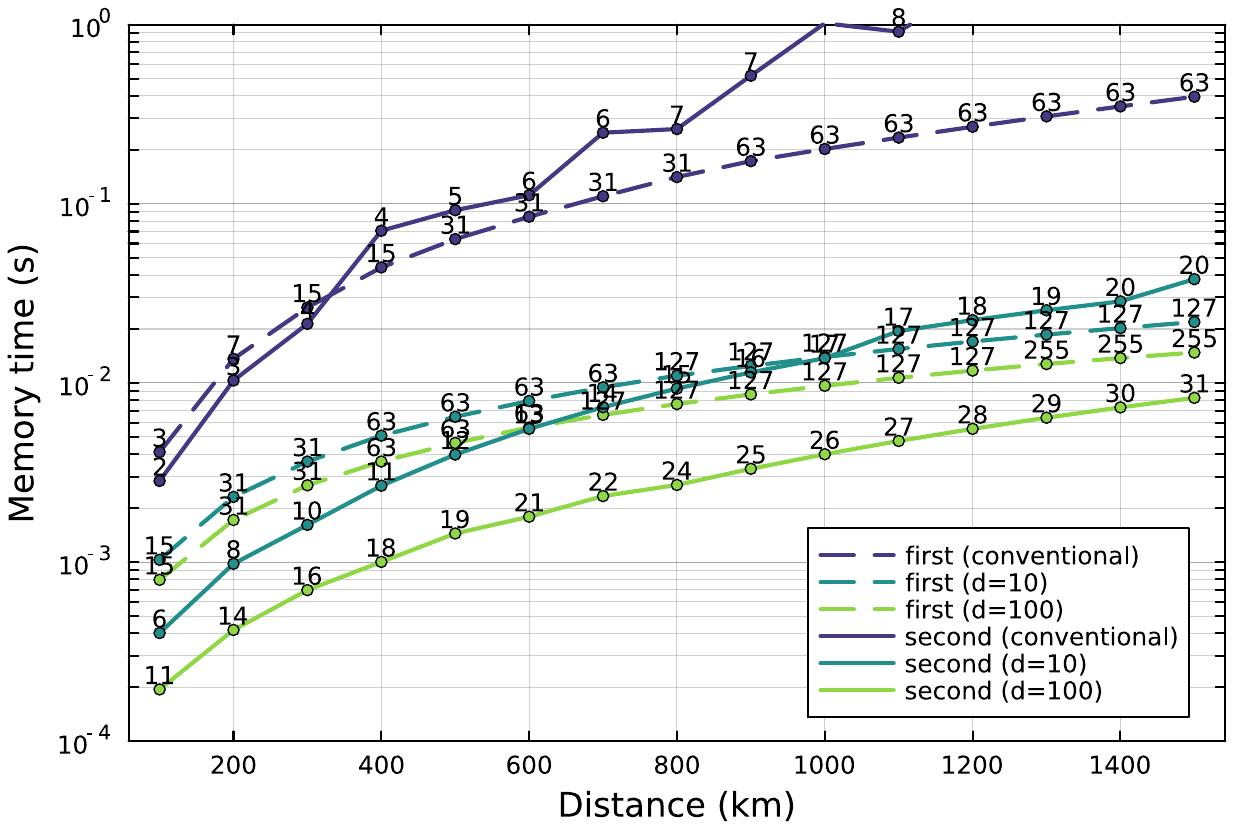} 
        \end{minipage}
    \caption{Distribution time per entangled pair of the protocols optimized with respect to the number of repeater nodes depending on the distance for (a) $\eta=0.95$ and (b) $\eta=0.99$, and memory time of the quantum memories required for the optimized protocol to operate for (c) $\eta=0.95$ and (d) $\eta=0.99$, respectively. The standard fusion gates and entanglement swapping are used in the conventional protocols while the pairwise fusion gates without ancilla photons and boosted entanglement swapping are used in the protocols with qudits of dimension $d=10$ or $d=100$. The number next to each plot represents the number of the repeater nodes used in each optimized protocol.}
    \label{fig:numerical}
\end{figure}

We perform numerical simulations of the distribution time and required memory time of the first- and second-generation quantum repeater protocols.
We let $\eta=\eta_\text{d} = \eta_\text{c}$ for simplicity and consider the case of $\eta = 0.95$ and $\eta = 0.99$.
In the conventional protocols, the standard fusion gate and entanglement swapping are used, where $P_s = \eta_\text{d}^2 \eta_\text{c}^2 p_\text{f}$ and $p_\text{f}=1/2$.
In our protocols, the pairwise fusion gates and boosted entanglement swapping are used, where $P_s = \eta_\text{d}^4 \eta_\text{c}^4 p_\text{f}$, and we assume that the pairwise fusion gates are implemented by using the linear-optical circuit without ancilla photons, that is, $p_\text{f} = 1-d^{-1}$ for qudits of dimension $d$.
We then consider the cases of $d=10$ and $d=100$.

The results of the numerical simulation are summarized in Fig~\ref{fig:numerical}.
For either $\eta$, our protocols with $d=10$ and $d=100$ outperform the corresponding conventional protocols in any case, which means that the use of the pairwise fusion gates and boosted entanglement swapping is beneficial.
In the first-generation protocols, the protocols with $d=10$ and $d=100$ are much better than the standard protocol in terms of the both distribution time and memory time while the difference between $d=10$ and $d=100$ is relatively small.
It seems to be consistent with the asymptotic behavior of $T^*_\text{first}$ with respect to $L_\text{tot}$, shown in Eqs.~\eqref{eq:constant}, \eqref{eq:upperbound}, and \eqref{eq:lowerbound}, because the values of $\alpha=3/(4P_\text{s})$ in this numerical simulation are about
$1.84, 1.26, 1.14, 1.56, 0.90, 0.82$ for $(\eta, d)=(0.95,2),(0.95,10),(0.95,100),(0.99,2),(0.99,10),(0.99,100)$, where $d=2$ corresponds to the use of the standard entanglement swapping in conventional protocols.
In contrast, the performances of $d=10$ and $d=100$ in the second-generation protocols are quite different for $\eta = 0.99$, which reflects a strong dependency on the success probability $P_\text{s}$ of the entanglement swapping.
(The increase of $P_\text{s}$ from $d=10$ to $d=100$ is relatively small in the case of $\eta = 0.95$. It would explain why the performance difference between the cases of $d=10$ and $d=100$ is not significant for $\eta = 0.95$.)
In comparison between the first- and second-generation protocols,
the second-generation protocols can outperform the corresponding first-generation protocols in terms of the distribution time only when $\eta=0.99$ and $d=100$, and the distance is shorter than \SI{300}{km}.
In the other cases, the first-generation protocols outperform the second-generation protocols.
On the other hand, the second-generation protocols require the memory time less than one required in the first-generation protocols for a wider range of distances in the case of $\eta=0.99$.
On the second-generation protocols, the success probability of the entire protocol becomes exponentially small with respect to the number of the repeater nodes.
Thus, the optimized number of the repeater nodes in the second-generation protocols tends to be much smaller than one in the first-generation protocols.
If we restrict the maximum number of the repeater nodes, which may be required to suppress the accumulation of operational errors on the entanglement swappings or the experimental cost to place additional repeater nodes, the performance curves in Fig.~\ref{fig:numerical} become straight lines rather than curves for large distances in which the number of the repeater nodes reaches the maximum.
Then, the performance of the first-generation protocols becomes worse and close to one of the second-generation protocols although the second-generation protocols rarely outperform the first-generation protocols even in this case.
For the second-generation protocols to outperform the first-generation protocols in reasonable experimental parameters, it would be necessary to use quantum error-correcting codes as in~\cite{Jiang_Phys.Rev.A2009,Munro_Nat.Photonics2010,Li_NewJ.Phys.2013}, which is left for future work.
\bibliography{hdbell.bib}
\makeatletter\@input{hdbell_aux.tex}\makeatother